\documentclass[11pt,a4paper]{article}
\usepackage{jheppub}
\usepackage{upquote} 

\usepackage{placeins}
\usepackage{tabularx,booktabs}
\usepackage[small]{caption}
\usepackage{lscape}
\newcolumntype{M}[1]{>{\centering\arraybackslash}m{#1}}
\newcommand{\vp}{{\vec p}}

\newcommand{\gtni}{\widetilde{u}}
\newcommand{\ut}{\widetilde{u}}
\newcommand{\vt}{\widetilde{v}}
\newcommand{\vq}{\vec{q}}

\newcommand{\re}{{\rm e}}
\newcommand{\rd}{{\rm d}}
\renewcommand{\d}{\partial}

\newcommand{\dmu}{\partial_\mu}
\allowdisplaybreaks 

\title{\texttt{Phi4tools}: Compilation of Feynman diagrams for Landau-Ginzburg-Wilson theories}

\author[a,b]{Giacomo Sberveglieri}
\author[c]{Gabriele Spada}

\affiliation[a]{SISSA and INFN, Via Bonomea 265, I-34136 Trieste, Italy}
\affiliation[b]{Albert Einstein Center for Fundamental Physics,
Institute for Theoretical Physics, University of Bern,
Sidlerstrasse 5, CH-3012 Bern, Switzerland}
\affiliation[c]{Pitaevskii BEC Center, CNR-INO and Dipartimento di Fisica, Universit\`a di Trento, Via Sommarive 14, I-38123 Povo, Trento, Italy}

\emailAdd{sberveglieri@itp.unibe.ch}
\emailAdd{gabriele.spada@ino.cnr.it}

\abstract{
    Scalar field theories with quartic interactions are of central interest in the study of second-order phase transitions.
    For three-dimensional theories, numerous studies make use of the fixed-dimensional perturbative computation of
    [\emph{B. Nickel, D. Meiron, and G. Baker Jr, Compilation of 2-pt and 4-pt graphs for continuous spin model, University of Guelph report (1977)}], unfortunately left unpublished.
    We independently verify the results of Nickel et al., and we extend the computation to the eighth order in the coupling constant.
    The results of our calculations, together with the tools developed, are made available in \texttt{Phi4tools}, a user-friendly package that allows displaying the information about the individual Feynman diagrams, including the numerical values
    for the diagrams 
    for zero, two, and four-point functions. We also provide the perturbative series up to order eight for the renormalization-group functions for the $O(N)$ and cubic anisotropic models.
}

\begin{document}

\maketitle

\section{Introduction}
\label{sec:intro}

High-order perturbative computations in quantum field theories are challenging because of the rapid growth in the number of diagrams with the perturbative order as well as in the complexity of the multi-dimensional integration.
The availability of automatic tools for the generation and simplification of Feynman diagrams mitigates the risk of errors, but users are still required to have the technical knowledge and the computational time for obtaining some results that might have already been computed by others.
This obstacle has an obvious impact on research, driving resources away from physics questions or discouraging them in the first place.
Conversely, when the results are shared in their raw data form, they facilitate collaboration and encourage further scientific exploration. This was the case for the multi-loop results for the scalar field theory with quartic interaction reported by Nickel et~al.~in the unpublished ref.~\cite{Nickel:1977gu}, and originally used in refs.~\cite{Baker:1976ff,Baker:1977hp}, that were subsequently exploited in several other works. A non-exhaustive list includes the study of the $O(N)$-symmetric model~\cite{LeGuillou:1977rjt,Guida:1998bx,Calabrese:2002qi}, the $N$-component model with cubic anisotropy~\cite{Carmona:1999rm}, randomly dilute spin models~\cite{Pelissetto:2000ra,Calabrese:2002sz}, self-avoiding walks on a cubic lattice \cite{Pelissetto:2007tx}, the shift in the Bose-Einstein condensation temperature~\cite{Kastening:2003iu}, the finite temperature chiral transition in QCD \cite{Basile:2004wa},
and frustrated spin systems \cite{Calabrese:2004at}.
The study of such a wide variety of models within fixed-dimensional field theory was accomplished by
combining the values of the Feynman diagrams shared by Nickel et al.~
with the corresponding symmetry factors obtained from the tensorial structure of the interaction vertices,
obviating the necessity of time-consuming multi-loop computations.
In the same spirit, with the hope of avoiding long calculations for others,
we share in this work the individual results for the Feynman diagrams of the Landau-Ginzburg-Wilson theory,
together with the tools we developed for the computations and for handling the data in ref.~\cite{Sberveglieri:2020eko}.
Our results include the values of the diagrams in three dimensions for the zero, two, and four-point functions with up to eight vertices and with zero external momenta, going beyond the six-loop results of ref.~\cite{Nickel:1977gu}.
The $N$-component Landau-Ginzburg-Wilson theory is defined by the Hamiltonian
\begin{equation} \label{eq:LGWHamiltonian}
    \mathcal{H} = \int d^dx \left[
    \frac12 \left( \dmu \phi_i \right)^2
    + \frac12 m_0^2 \phi_i^2
    + \frac{1}{4!} \lambda_{i j k l} \phi_i \phi_j \phi_k \phi_l
    \right]
    \,,
\end{equation}
describing $N$ scalar fields $\phi_i$, $(i = 1, \dots, N)$ interacting via a generic quartic coupling $\lambda_{i j k l}$.\footnote{Summation over repeated indices is implied.}
Perturbative calculations in the coupling $\lambda_{ijkl}$ can be factorized into the computation of symmetry factors, multiplied by the integrals of scalar propagators that depend only on the topology of the diagrams, allowing one to reuse the expensive and tedious numerical computations for the study of many different theories.\\

This paper accompanies the Wolfram Mathematica \cite{Mathematica} package \texttt{Phi4tools} \cite{phi4tools:2023zen} and describes the computational strategies adopted. \texttt{Phi4tools} is an intuitive interface for visualizing, simplifying, and manipulating Feynman diagrams for the Landau-Ginzburg-Wilson theory. In more detail, it allows displaying detailed information about the diagrams, including their Nickel indices, the integrands, and numerical results. It also provides the symmetry factors for the $O(N)$-symmetric model and the $N$-component model with cubic anisotropy. Diagrams with cubic vertices, described by a symmetry-broken version of \eqref{eq:LGWHamiltonian} with an added cubic interaction $\eta_{ijk}\phi_i \phi_j \phi_k$, are also implemented but not computed.\\

\texttt{Phi4tools} is fundamentally different from other existing software, its main focus being on providing user-friendly access to pre-computed quantities in scalar field theories, while, at the same time, packing some general tools for the manipulation and computation of other diagrams, possibly extending its own scope.
It is not meant to substitute other widely used software for Feynman diagrams, usually aimed at performing one specific task as efficiently as possible and targeting general theories. Loosely speaking, this software can be divided into two groups: One aimed at the generation of the diagrams (topologies, symmetry factors, and symbolic expressions), which includes programs such as \texttt{QGRAF} \cite{Nogueira:1991ex}, \texttt{FORM} \cite{Ruijl:2017dtg}, and \texttt{feyngen} \cite{Borinsky:2014xwa}. The other aimed at the (semi)automatic reduction, simplification and evaluation of these symbolic expressions, which includes \texttt{LiteRed} \cite{Lee:2013mka}, \texttt{Reduze} \cite{Studerus:2009ye}, \texttt{FIRE} \cite{Smirnov:2008iw,Smirnov:2019qkx}, \texttt{Kira} \cite{Maierhofer:2017gsa,Klappert:2020nbg}, and \texttt{tapir} \cite{Gerlach:2022qnc} among the others. Some of the above programs can be used to generate input files for \texttt{Phi4tools} or used in conjunction with it. 
We also mention the presence of the online database \texttt{Loopedia},\footnote{The database can be found at the following URL \url{https://loopedia.mpp.mpg.de/}.} which aims at collecting the $\epsilon$-expansion results found in the literature \cite{Bogner:2017xhp}, and the impressive seven-loop $\epsilon$-expansion computation of ref.~\cite{Schnetz:2022nsc}. \\

The rest of the paper is organized as follows:
in section \ref{sec:numerical} we present the details of the computations,
while in section \ref{sec:resummation} we utilize the results to generate the perturbative series for the RG functions in the $O(N)$ and cubic-symmetric models up to the eighth order.
We conclude in section \ref{sec:conclusions}. Five appendices complete the paper. In appendix \ref{app:Conventions}, we present the conventions and the normalizations adopted.
In appendix \ref{app:eff_vertices}, we report the analytical expression for the effective vertices in three dimensions.
In appendix \ref{app:paclet}, we provide a brief introduction to the \texttt{Phi4tools} paclet itself, with instructions on how to install it and some basic examples.
In appendices \ref{app:coeff_O(N)} and \ref{app:CoeffCub}, we report the explicit form of the perturbative series up to the eighth order in the coupling for the $\beta$-function, $\eta$, and $\nu$ in the $O(N)$ model and $\beta$-function, $\eta$, and $\eta_2$ for the $N$-component model with cubic anisotropy, respectively.

\section{Fixed-dimensional perturbative computations}
\label{sec:numerical}

The evaluation of Feynman diagrams in super-renormalizable quantum field theories poses some technical challenges that must be addressed in order to push the computation to high orders. These challenges are two-fold: Firstly, as the perturbative order increases, the complexity escalates due to the proliferation of diagram topologies, each requiring the calculation of symmetry factors. Secondly, the dimensionality of each integral grows with the number of loops, posing significant computational challenges in achieving precise results for the perturbative series.
To address this issue, a possible strategy is to reduce the dimension of the integration space by performing part of the integration analytically \cite{Nickel:1978ds}. An effective approach involves working in momentum space, identifying simple subdiagrams within the Feynman diagrams, and replacing them with their analytical expressions \cite{Guida:2005bc}.
This is the central operation that we utilize to compute the perturbative series up to the eighth order. However, there are several other measures that turned out to greatly reduce the computational cost and improve the accuracy of our results. 
Our method follows the one reported in ref.~\cite{Guida:2005bc}, with only slight differences: we have opted to separate each step of the computation and to use different tools at each step. We have also introduced some new effective vertices (analytic and numeric) to further simplify some topologies, but, on the other hand, we didn't implement some of their more sophisticated approaches for the parametrization of the amplitudes. The final complexity of the computation is very similar.
In the following, we summarize the main steps of our computation, starting from the choice of the renormalization scheme.

\subsection{Renormalization scheme}

In $d < 4$, scalar field theories with quartic interactions described by the Hamiltonian eq.~(\ref{eq:LGWHamiltonian}) are superrenomalizable and the divergences can be absorbed with just a shift in the mass parameter and a vacuum-energy renormalization constant that we neglect here. Indeed, after the renormalization of the tadpole and the sunset diagrams, the theory is finite and, therefore, amenable to fixed-dimension computations.
From a practical point of view, it is convenient to get rid of the divergences
by performing a zero-momentum subtraction of the divergent integrands \cite{Zimmermann:1969jj,Guida:2005bc,Sberveglieri:2020eko}.
In this scheme
 , labeled with the subscript $I$ as in refs.~\cite{Guida:2005bc,Sberveglieri:2020eko}, the mass counterterm $\delta m^2_I$ completely cancels the one-loop tadpole diagram and removes the divergence coming from the sunset diagram in such a way that the regularized sunset diagram is exactly zero at $p=0$, namely
\begin{align}
     m^2_I & = m^2_0 + \delta m^2_I\,, \\
    \delta m^2_I & = - \left( ~\raisebox{-0.4cm}{\includegraphics[height=1cm]{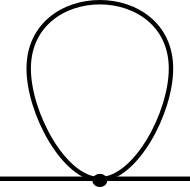}} + \raisebox{-0.4cm}{\includegraphics[height=1cm]{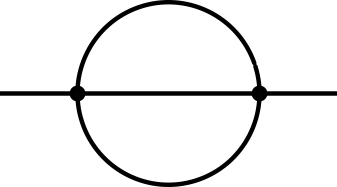}}{}^{\hspace*{-10pt} p=0} ~\right) \,.
    \label{eq:dm2I}
\end{align}
The renormalization procedure then amounts to setting the tadpole subdiagrams to zero and to substituting the sunset subdiagrams with their regularized versions.
The above prescription makes the theory finite for any generic coupling $\lambda_{i j k l}$ in $d<4$ and provides a convenient setup for the numerical evaluation of high-order diagrams in fixed dimensions \cite{Baker:1976ff,Baker:1977hp}.\footnote{The renormalization scheme adopted in \cite{Nickel:1977gu} is different: the counterterm subtracts the whole two-point subdiagrams at zero momenta instead of just the divergent tadpole and sunset diagrams.
}

\subsection{Drawing and labeling the Feynman diagrams}
\label{sec:labeling}

After having chosen the renormalization scheme, the first step to compute the perturbative series consists in the generation of all the Feynman diagrams at each given order, together with their multiplicity factors. We leverage the already available \texttt{feyngen} program \cite{Borinsky:2014xwa} to generate all the one-particle-irreducible (1PI) diagrams for the zero, two, and four-point functions with up to nine quartic vertices. The choice of a renormalization scheme in which the regularized tadpole is zero greatly reduces the number of Feynman diagrams that one has to compute since all the diagrams involving tadpoles vanish. These diagrams are completely omitted in the \texttt{Phi4tools} package. We also generate the diagrams for the theory with both quartic and cubic interactions with up to eight total vertices using a simple program that, starting from the diagrams with only quartic vertices, repeatedly removes propagators, accounts for the proper correction to the symmetry factor, and finally collects the resulting diagrams based on their topology.\\

\begin{figure}[ht]
    \begin{center}
    \begin{tabular}{  M{4.5cm} M{4.5cm} M{4.5cm} }
        \includegraphics[scale=0.35]{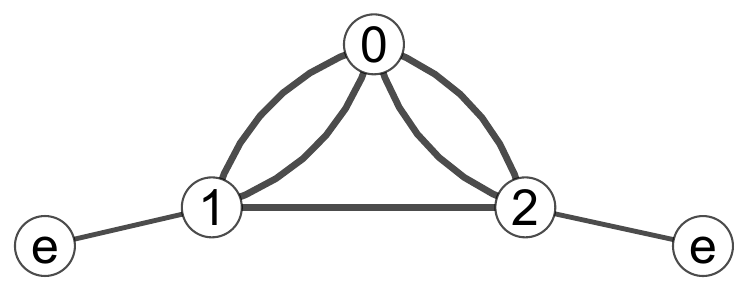} & \includegraphics[scale=0.35]{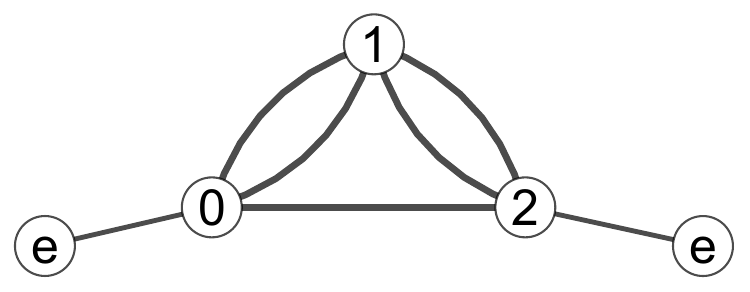}  & \includegraphics[scale=0.35]{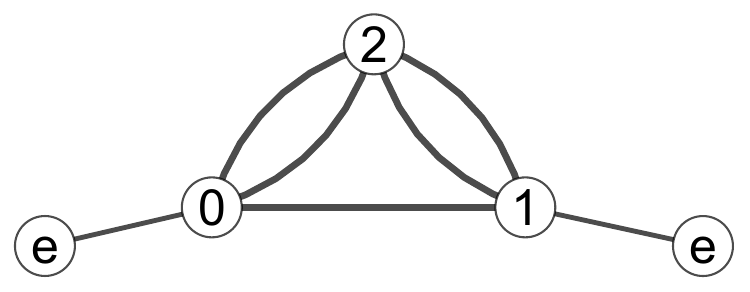}
    \end{tabular}
    \begin{equation*}\label{Nick2}
        1122|\re 2|\re | \hspace{1.5cm} > \hspace{1.5cm} \re 112|22|\re | \hspace{1.5cm}< \hspace{1.5cm} \re 122|\re 22||\,.
    \end{equation*}
    \caption{Different enumerations and their corresponding sequences. The first one is the biggest of the three since its first digit is 1 and  $1>\re$ and the central is smaller than the last since $\re11<\re12$. The central one is, hence, the smallest one, the minimal graph descriptor.}
    \label{Nickel_example}
    \end{center}
\end{figure}

To systematize and categorize the diagrams, we adopt
the commonly used \emph{Nickel index} \cite{Nagle:1966jm,Nickel:1977gu,Batkovich:2014bla} to label them.
For completeness, we briefly outline in what follows the labeling algorithm, explaining the process of assigning an index to a given graph and how to interpret it. Consider an arbitrary undirected connected graph with $n$ internal vertices, already labeled from $0$ to $n-1$, and with some external vertices all labeled  ``e". The Nickel index for this labeled graph $\mathcal{G}_\textnormal{L}$ is the sequence constructed in the following way
\begin{equation}\label{eq:Nick1}
\mathfrak{N}\left(\mathcal{G}_\textnormal{L}\right) = c(0)|c(1)|\dots|c(n-1)|\,,
\end{equation}
where $c(i)$ is the sequence of all the vertices connected to the vertex $i$ whose label is $j \ge i$, repeated in case of multiple edges, and ordered in ascending order with the convention that ``e" goes first in the order, i.e. $\re<0<1<...< n-1$.
In this way, the sequence~\eqref{eq:Nick1} corresponds exactly to one graph, which can be directly reconstructed from the sequence.
However, the opposite is not yet true since the string~\eqref{eq:Nick1} depends on the way we enumerated the vertices.
In order to overcome this ambiguity, we first establish a way to order different sequences: the strings obtained as above are converted to a numeric field, interpreting each sequence as a number with radix $n+2$ with the following order to its digits $\re<``|"<0<1<...< n-1$. The correct labeling of the vertices is then identified as the one that corresponds to the sequence with the smallest number, the minimal graph descriptor. 
For example, in figure~\ref{Nickel_example} we show three possible labelings of the same diagram. 
The central label is the minimal one, hence the correct Nickel index associated with the diagram.

With a simple program, we assigned the Nickel indices to the Feynman diagrams at each order and ordered them according to their graph descriptor.
In ref.~\cite{phi4tools:2023zen} we provide the text files containing the list of the Feynman diagrams at each order, consisting in their Nickel index, edge list (where the name of the vertices is already the one used by the Nickel index), and weight factor. See appendix \ref{app:Conventions} for the definition of weight factor and the conventions used in this work. The \texttt{Phi4tools} package allows easy access to the data, providing a convenient interface for the visualization of the graphs and their Nickel indices, see appendix \ref{app:paclet}.

\subsection{Symmetry factors: $O(N)$ model and cubic anisotropy}

Once the topologies of the Feynman diagrams are known, it is not difficult to compute their corresponding symmetry factors, given a tensorial structure for the quartic coupling $\lambda_{ijkl}$ in the Hamiltonian of eq.~(\ref{eq:LGWHamiltonian}).
In this work, we do this for the $N$-component cubic-symmetric theory, in which we have
\begin{equation}\label{eq:TensLCub}
    \lambda_{ijkl} = \frac{u_0}{3}(\delta_{ij}\delta_{kl}+\delta_{ik}\delta_{jl}+\delta_{il}\delta_{jk})+v_0 \ \delta_{ij}\delta_{ik}\delta_{il}\,.
\end{equation}
The symmety factors for the $O(N)$-symmetric models can be readily deduced by turning off the coupling $v_0$.
With a simple program that assigns tensor factors to the vertices and performs contractions, we computed the symmetry factors for these models for the zero, two, and four-point functions up to order eight.
In ref.~\cite{phi4tools:2023zen} we provide those lists in text files. We suggest \texttt{Phi4tools} package to quickly navigate through them. We refer once again the reader to the appendix \ref{app:Conventions} for normalization and conventions and to appendix \ref{app:paclet} for a quick introduction to the package.

\subsection{Substitutions of the effective vertices}
\label{sec:substitutions}

The number of loops of a diagram is given by $l = v_4 + v_3/2 - e/2 + 1$, where $v_4$ and $v_3$ are the number of quartic and cubic vertices respectively and $e$ is the number of external lines. Employing spherical coordinates and leveraging the symmetries of the integrands, the dimension $D$ of the integration space (in $d=3$ spatial dimensions) is $D=1$ for $l=1$ and $D=3l-3$ for $l>1$. Directly expressing the integrands in momentum space and performing the integrations would result in very demanding computations. To illustrate the scale of this task, consider, for example, the two-point function with eight quartic vertices, where each of the $1622$ non-zero 1PI diagrams would require a $21$-dimensional numerical integration.
Nevertheless, it is possible to substantially reduce the complexity of the integration by substituting analytically known subdiagrams, as done long ago by Baker,
Nickel, Green, and Meiron in refs.~\cite{Baker:1976ff,Baker:1977hp}, where they used the analytically known expressions for the one-loop subdiagrams \cite{Melrose:1965re}  to compute the six-loop $\beta$-function.
These substitutions can be performed directly at the diagrammatic level, before writing the explicit form of the integrands. The approach involves identifying the cycles corresponding to known subdiagrams within the graphs and then replacing the propagators that constitute these cycles with complex vertices. The complex vertices are linked to the graph by new edges that, differently from the propagators, do not contribute to the integrand and just represent a bookkeeping device for the substituted topology.
After the substitution, we get a simplified graph corresponding to an \emph{effective diagram} with a reduced number of loops $\ell$.
By applying the same procedure to the sunset subdiagrams, we renormalize its divergent contribution, making all the diagrams finite\footnote{The zero-point diagrams up to order three are divergent as well, we set them to zero following the renormalization scheme presented in ref.~\cite{Sberveglieri:2020eko}.} (remember that the diagrams with tadpoles are already set to zero and removed from the diagram list).
We summarize below the substitutions that we have defined, showing the symbols we use to denote the effective vertices and the names of the functions to which they correspond.
To streamline this section, the analytical formulae for the effective vertices have been consolidated into appendix~\ref{app:eff_vertices}.
\begin{description}
\item[Renormalization of sunset subdiagrams.] We identify the \textit{sunset} subdiagrams and substitute them with analytical effective vertices.
\begin{center}
    \begin{tabular}{ |M{3.5cm}|M{3.5cm}|M{4cm}| }
        \hline
        Subdiagram & Effective vertex & Renormalized function \\
        \hline
        \includegraphics[scale=0.24]{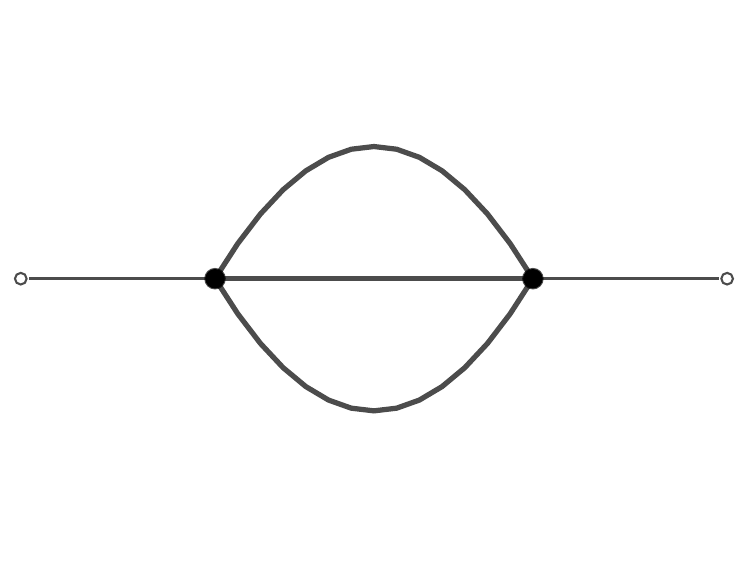} & \includegraphics[scale=0.23]{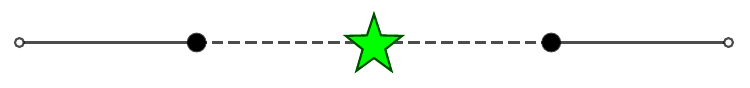}  & $\mathcal{S}(p)$ \\
        \hline
        \end{tabular}
\end{center}
\item[One-loop subdiagrams: bubbles, triangles and squares.] We identify three one-loop insertions, corresponding to cycles of length 2, 3, and 4, called respectively \textit{bubble}, \textit{triangle}, and \textit{square}, and substitute them with effective vertices that correspond to the analytic functions of the external momenta.
\begin{center}
        \begin{tabular}{ |M{3.5cm}|M{3.5cm}|M{4cm}| }
            \hline
            Subdiagram & Effective vertex & Analytic function \\
            \hline

            \includegraphics[scale=0.2]{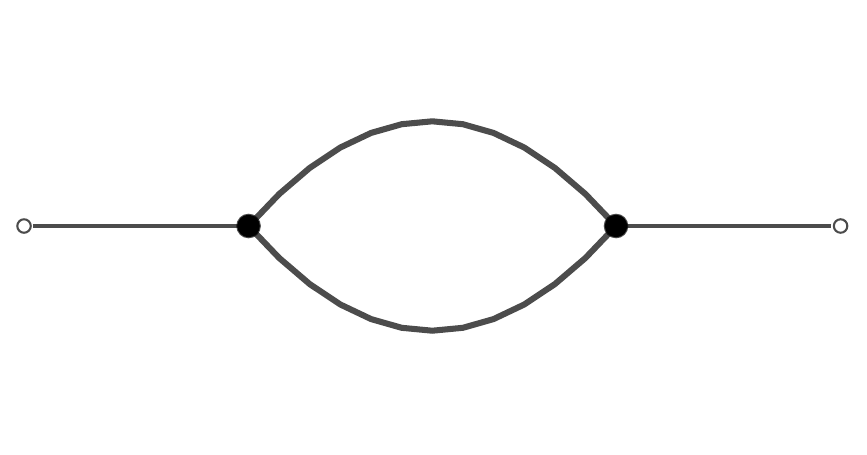} & \includegraphics[scale=0.22]{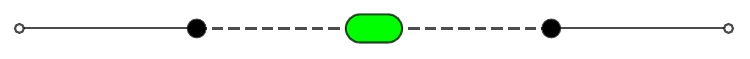}  & $\mathcal{B}(p)$ \\

            \includegraphics[scale=0.24]{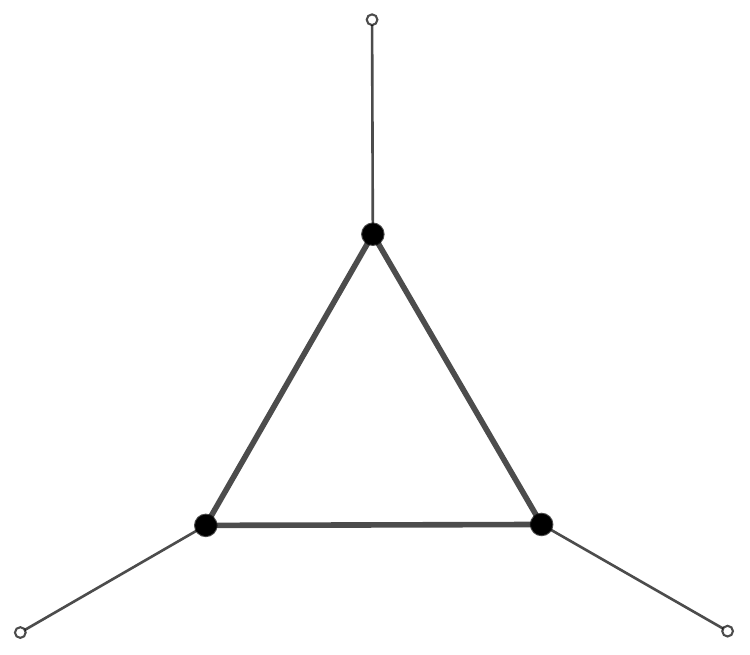} & \includegraphics[scale=0.23]{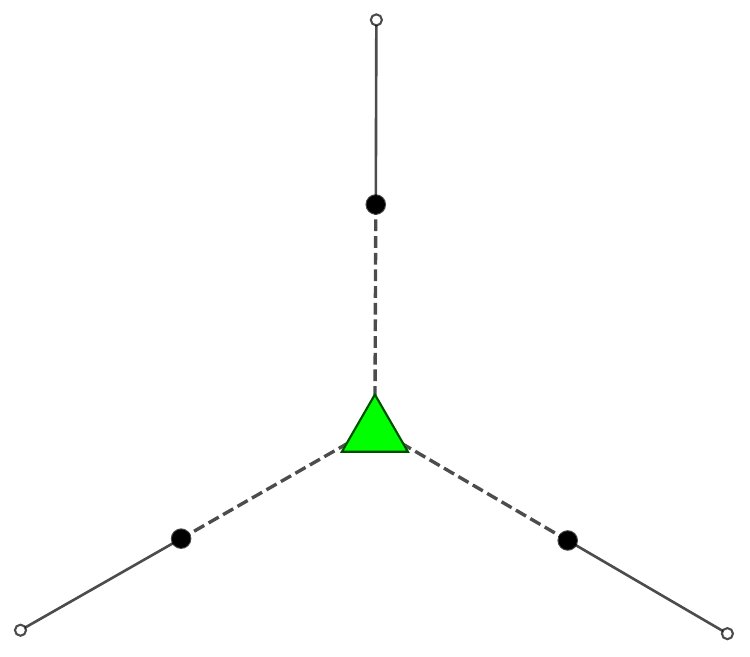}  & $\mathcal{T}(p_1,p_2,p_3)$ \\
            & & \\

            \includegraphics[scale=0.24]{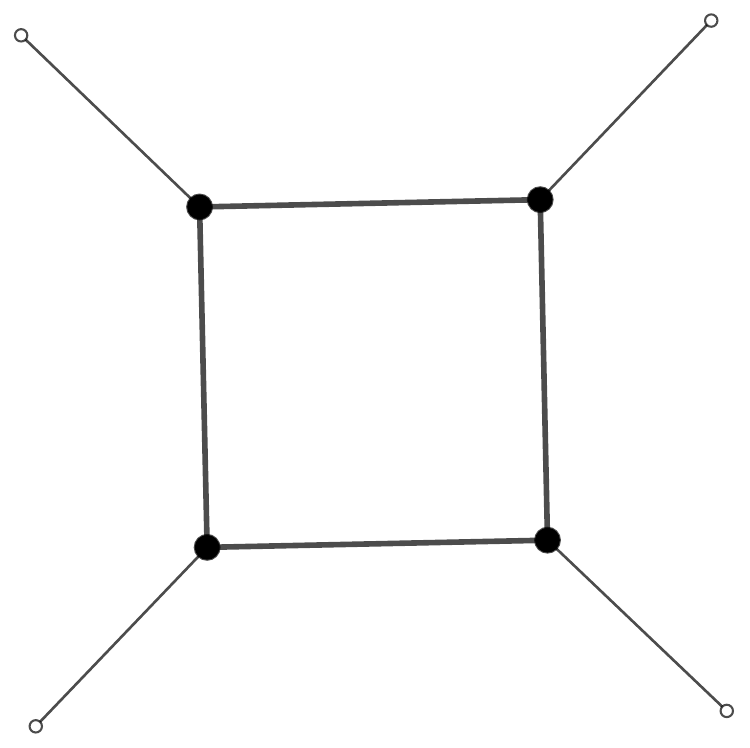} & \includegraphics[scale=0.23]{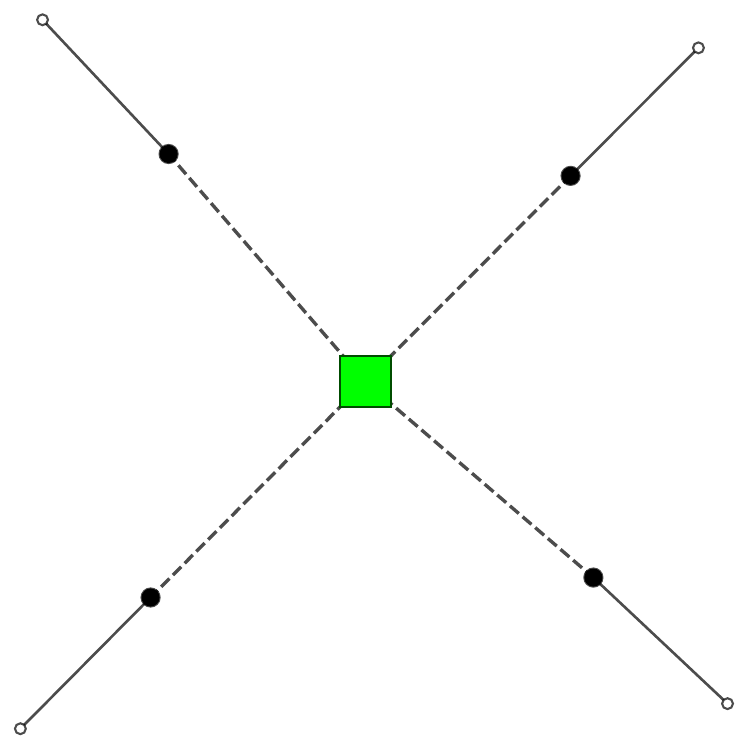}  & $\mathcal{Q}(p_1,p_2,p_3, p_4,$ $|\vp_1+\vp_2|,|\vp_2+\vp_3|)$ \\
            \hline
        \end{tabular}
\end{center}
\end{description}
Carrying on with this same philosophy, we identify other specific structures in the diagrams that are amenable for substitutions. These are momentum-independent subdiagrams, for which we know the analytic value, and subdiagrams depending on a single external momentum $\vec{p}$, that are substituted either with an analytic function or with a numeric approximation constructed from a tabulation of its values as a function of the magnitude $p \equiv |\vec{p}|$.
\begin{description}
\item[Momentum independent subdiagrams.] We identify two tadpole-like insertions that can be integrated analytically and we add new \emph{effective tadpoles} for them.
\begin{center}
    \begin{tabular}{ |M{3.5cm}|M{3.5cm}|M{4cm}| }
        \hline
        Subdiagram & Effective vertex & Analytic factor \\
        \hline
        \includegraphics[scale=0.2]{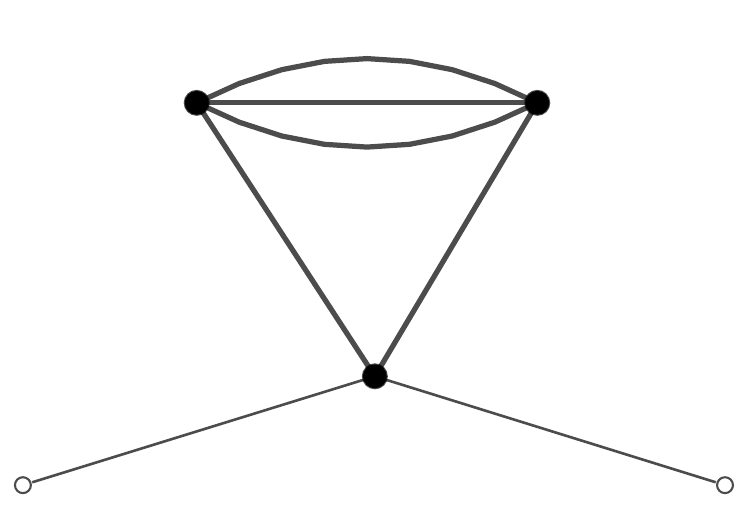} & \includegraphics[scale=0.22]{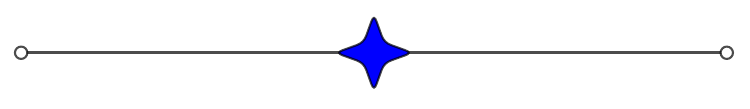}  & $t_{\mathcal{S}}$ \\

        \includegraphics[scale=0.2]{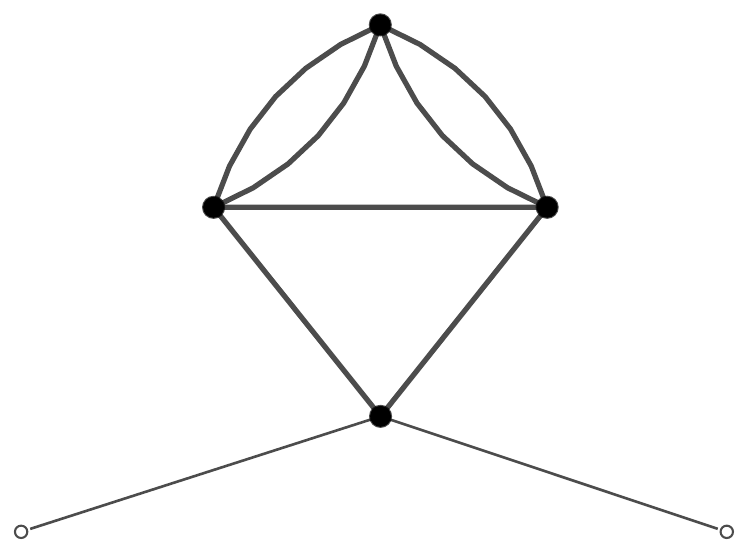} & \includegraphics[scale=0.22]{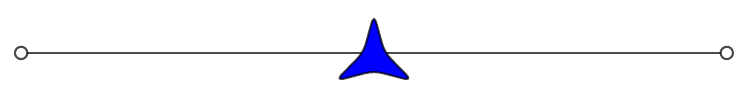}  & $t_{\mathcal{TB}}$ \\

        \hline
    \end{tabular}
\end{center}
\item[Numerical two-point subdiagrams.] We identify some multi-loop subdiagrams that depend just on one external momentum $p$ and substitute them with new effective vertices. We have constructed numeric functions for them, tabulating their value as a function of~$p$.
\begin{center}
    \begin{tabular}{ |M{3.5cm}|M{3.5cm}|M{4cm}| }
        \hline
        Subdiagram & Effective vertex & Numeric function \\  \hline
        \vspace{0.5cm}

        \includegraphics[scale=0.2]{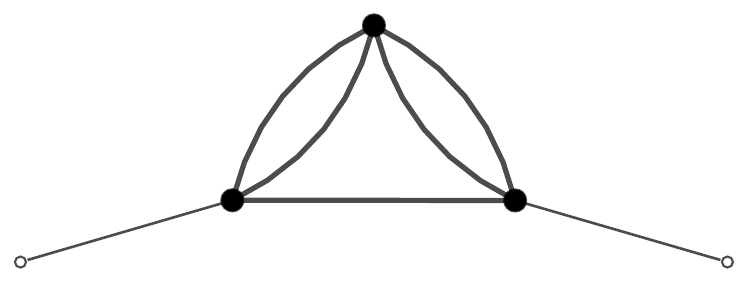} & \includegraphics[scale=0.22]{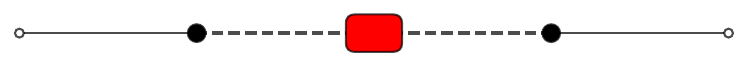}  & $\mathcal{T}_{bb}(p)$ \\
        \vspace{0.5cm}

        \includegraphics[scale=0.24]{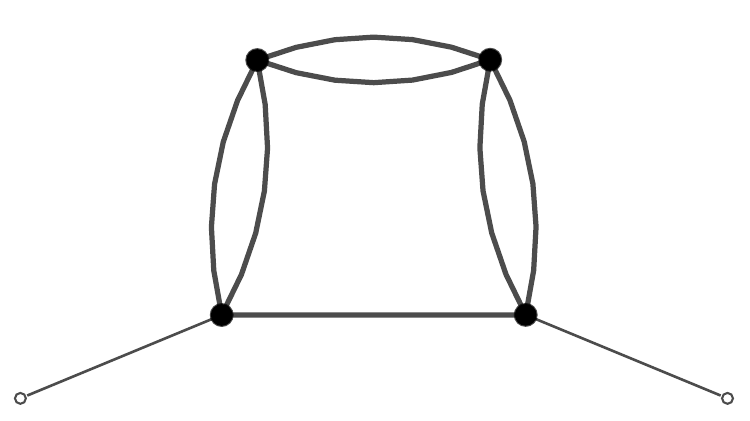} & \includegraphics[scale=0.23]{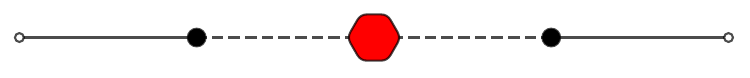}  & $\mathcal{Q}_{bbb}(p)$ \\
        \vspace{0.5cm}
        \includegraphics[scale=0.24]{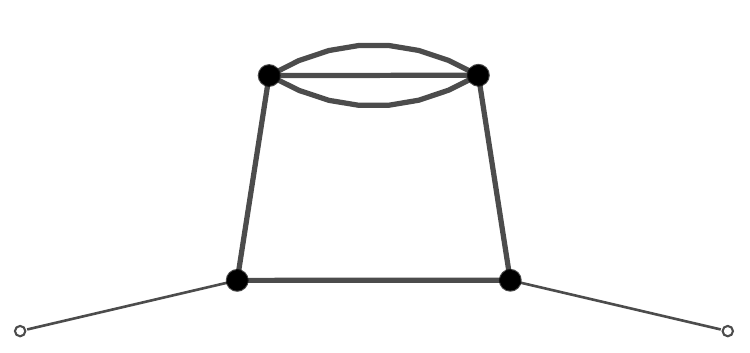} & \includegraphics[scale=0.23]{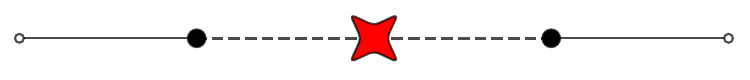}  & $\mathcal{Q}_s(p)$ \\
        \vspace{0.5cm}

        \includegraphics[scale=0.24]{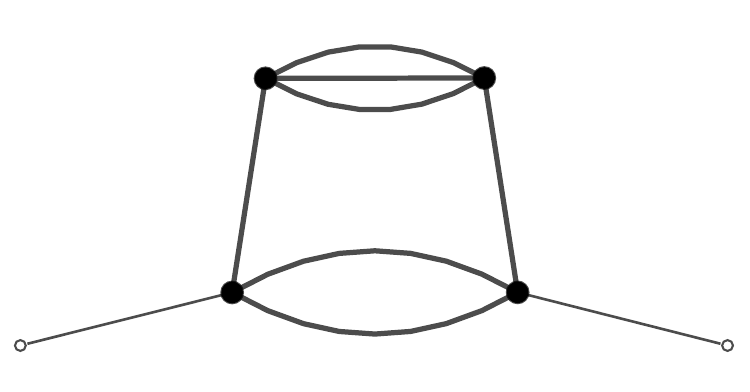} & \includegraphics[scale=0.23]{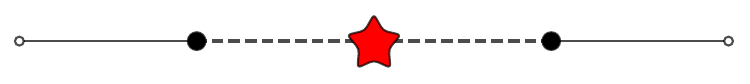}  & $\mathcal{Q}_{sb}(p)$\\

        \hline
    \end{tabular}
\end{center}
\item[Three and four-point subdiagrams with zero momentum flow through some legs.] We identify other two multi-loop subdiagrams that would in principle depend on more than one external momenta, but we focus on the combination where all except one are equal to zero. These substitutions affect the portions of the diagrams adjacent to the external legs. We have analytical expressions for some of the combinations, while we resort again to numerical tabulations for the others. We report these substitutions below, where we mark with a dotted line the external legs with zero momentum flow. The first substitution is analytical, the other two are numerical.
\end{description}
\begin{center}
\begin{tabular}{ |M{3.5cm}|M{3.5cm}|M{4cm}| }
    \hline
    Subdiagram & Effective vertex & Function \\
    \hline

    \includegraphics[scale=0.24]{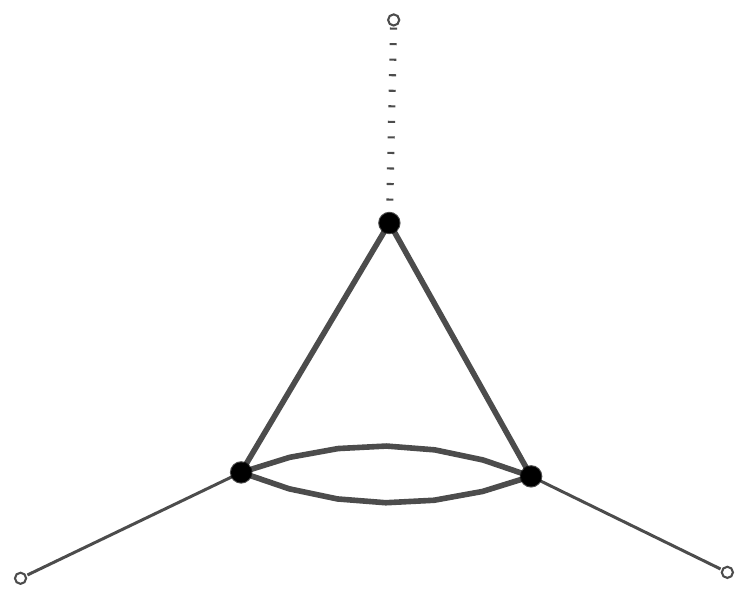} & \includegraphics[scale=0.23]{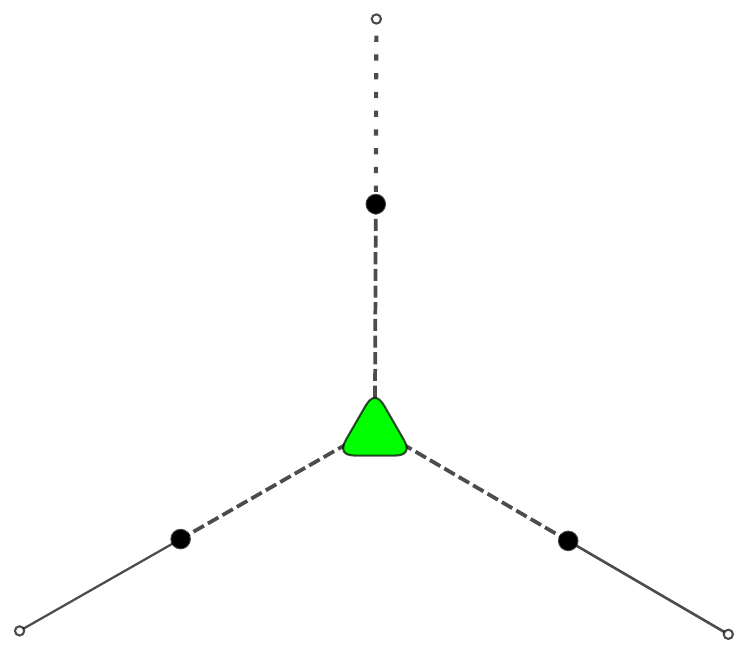}  & $\mathcal{T}_{bA}$($p$) \\

    \includegraphics[scale=0.24]{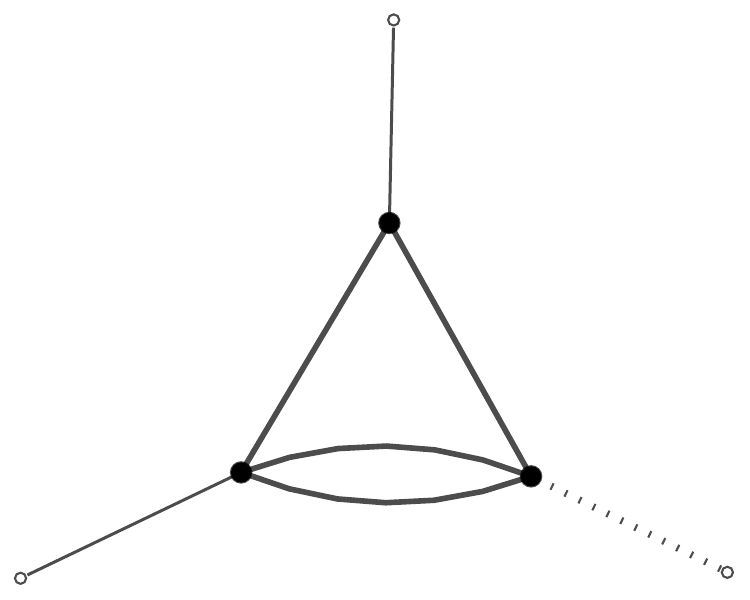} & \includegraphics[scale=0.23]{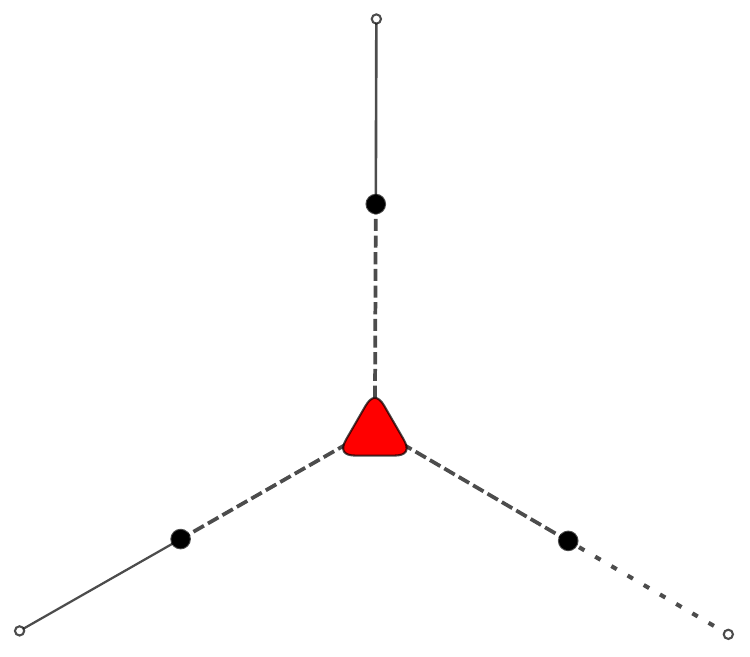}  & $\mathcal{T}_{bN}$($p$) \\

    \includegraphics[scale=0.15]{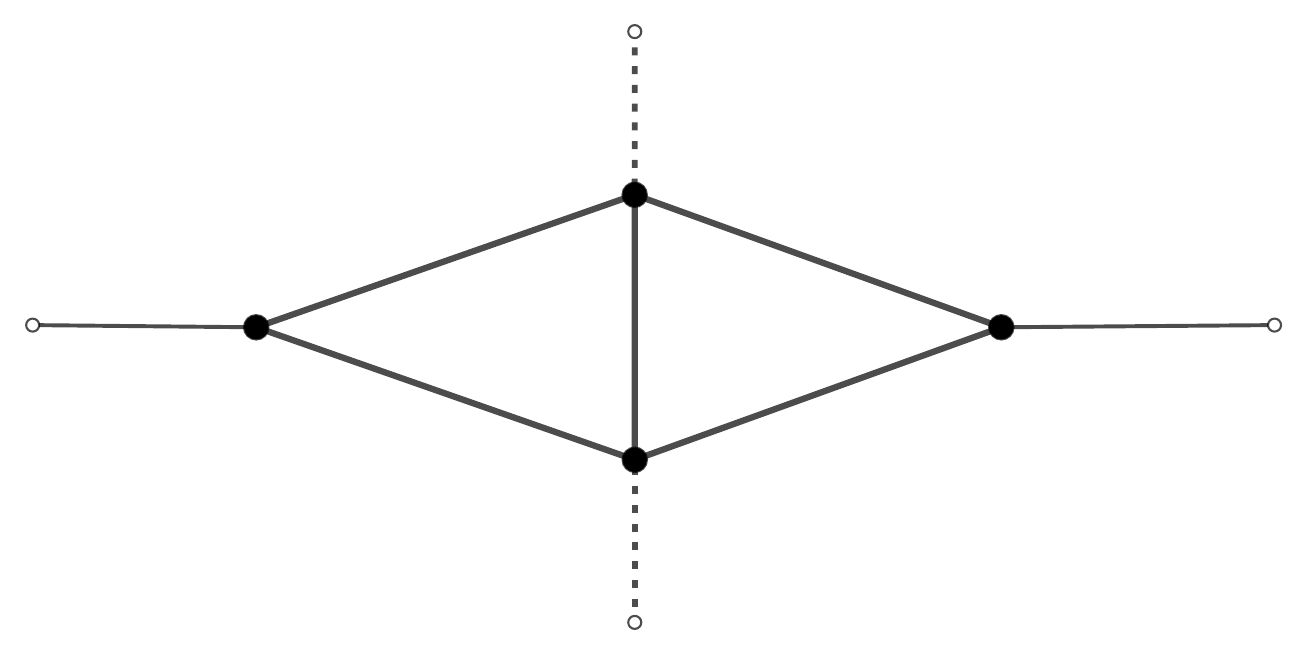} & \includegraphics[scale=0.21]{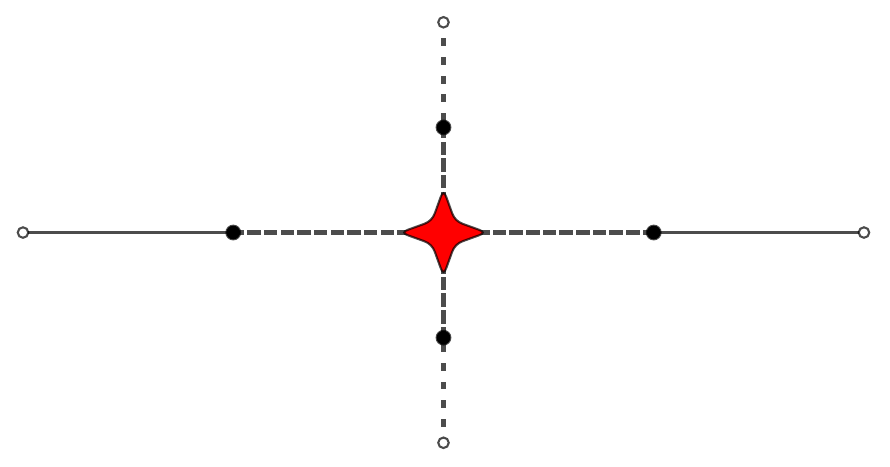}  & $\kappa$($p$)
     \\

    \hline
\end{tabular}
\end{center}

Clearly, for complex enough diagrams, there might be more than one possible set of substitutions that could translate into different numerical performances. In our implementation we choose to minimize the residual number of loops $\ell$ and, in case of parity, we prioritize effective diagrams with a higher number of analytic substitutions.
Nevertheless, it is not uncommon to encounter situations where equivalent effective diagrams exist. Therefore, we use an algorithm that allows us to favor specific substitutions over others through a set of weights. These weights are determined based on our empirical experience, noticing that on the one side, squares and triangles seem to produce better results since they reduce the number of propagators left in the integrand, but on the other side, they have a more complicated analytical form that depends on many scalar products that tend to produce more cumbersome integrands which are slower to evaluate.
In any case, the best combination of substitutions is diagram dependent and, for the few integrals in which the automatic choice of effective diagram does not produce precise enough results, we test the other equivalent parametrizations to find the one with the best performance.\\

\begin{table}[t]
\centering
\begin{tabular}{ |M{4cm}|M{4.6cm}|M{5cm}| }
    \hline
    Feynman diagram & Effective diagram & Effective integrand \\
    \hline
    \includegraphics[scale=0.25]{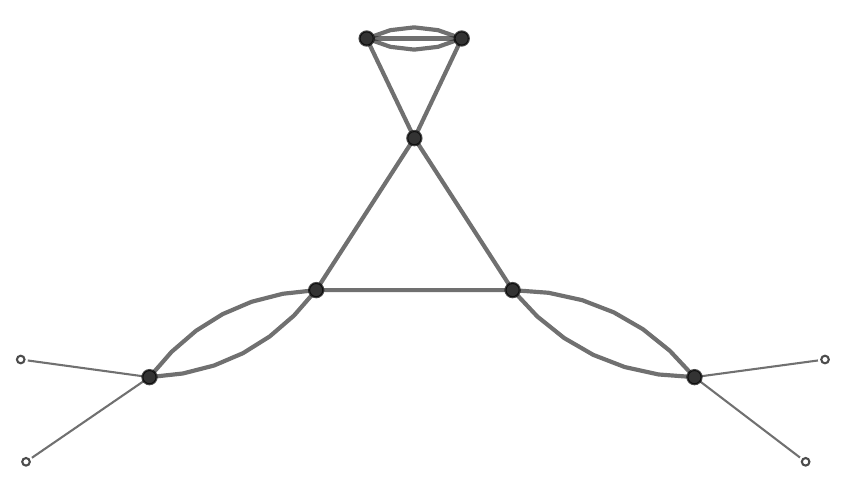} & \includegraphics[scale=0.235]{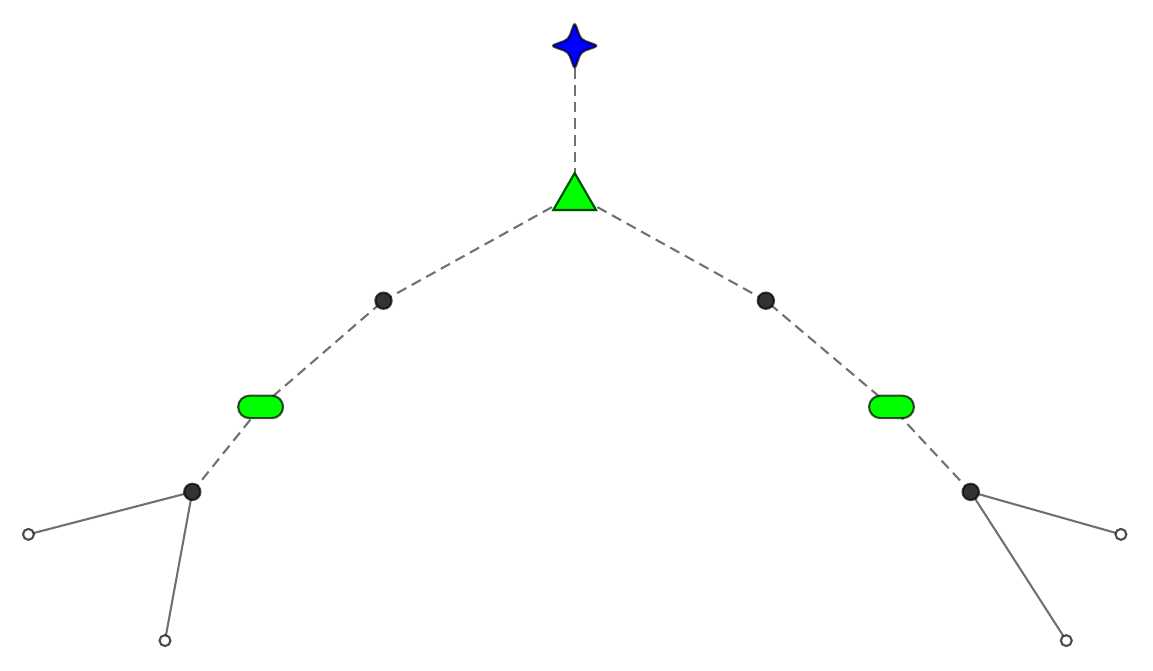}  & $\mathcal{B}(0)^2 \, t_{\mathcal{S}}$ $\mathcal{T}(0,0,0)$\\

    \includegraphics[scale=0.25]{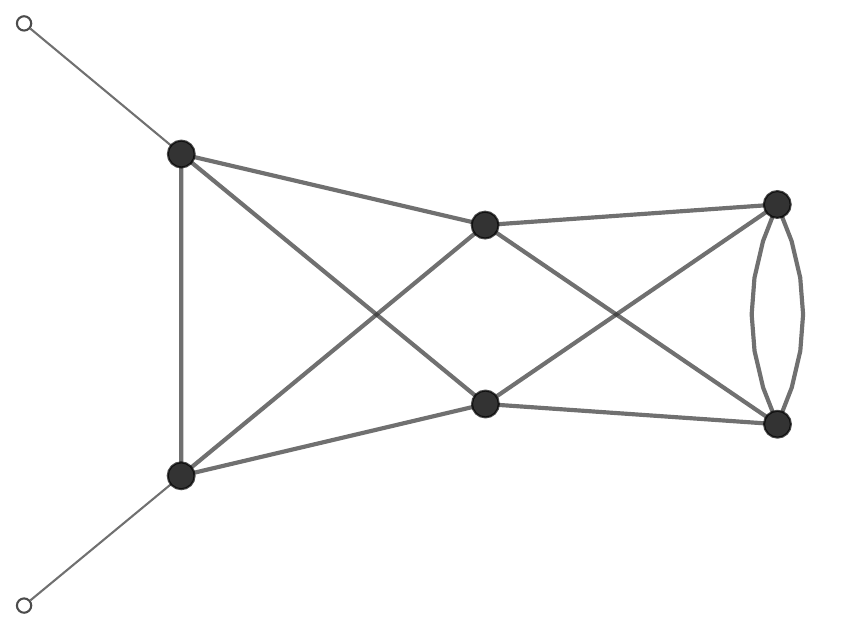} & \includegraphics[scale=0.25]{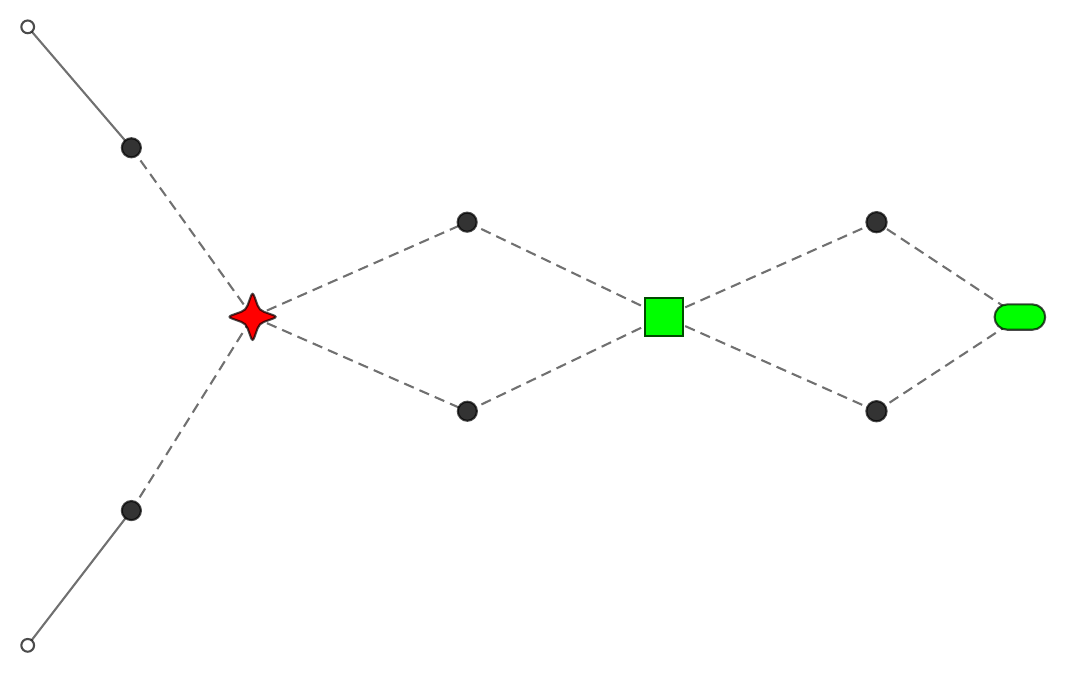}  & $\kappa(q_1) \mathcal{B}(q_2)$ $\mathcal{Q}(q1, q2, q1, q2,|\vq_1-\vq_2|,|\vq_1+\vq_2|)$\\

    \hline
\end{tabular}
\caption{Examples of substitutions for two Feynman diagrams. In the first row, we show a four-point diagram at order seven that results in a zero-loop effective diagram after the analytical substitutions of two bubbles, one triangle, and one complex tadpole. In the second row, we show a two-point diagram at order six that results in a two-loop effective diagram after the substitution of a bubble, a square, and a numerical two-point subdiagram. }
\label{tab:5}
\end{table}

As an example of how much these substitutions help in reducing the computational cost, we consider the example of the 1PI four-point function at zero external momenta $\Gamma^{(4)}(0)$ with five quartic vertices ($v_4=5$), consisting of 27 diagrams with four loops (corresponding to $D=9$ integrals) before the substitutions.
After the substitutions, we get 5 diagrams with zero residual loops (no integration left), 18 diagrams with one residual loop, and 4 diagrams with two residual loops. Therefore, we just have to perform 18 one-dimensional integrals and 4 three-dimensional integrals.
In table~\ref{tab:5} we show two examples of six-loop diagrams that, after the substitution of analytic and numeric effective vertices, result in a zero-loop effective diagram and a two-loop effective diagram.
With the \texttt{Phi4tools} package, it is possible to draw the Feynman diagrams before and after the insertion of the first three classes of effective vertices, e.g.~see figure \ref{fig:exampleInformationD} in appendix~\ref{app:paclet}.

\subsection{Momentum assignation: writing the integrands}

We are now ready to write the integrands associated with the effective diagrams.
It is quite straightforward to automatically implement the momentum assignation at every internal edge and then impose the conservation of total momenta at each (effective) vertex.
A generic diagram with $\ell$ effective loops then corresponds to an integral over the $\ell$ internal momenta of a function of the form $f(\vec{ q}_1,...,\vq_\ell)$.
Additionally, one can make use of the spherical symmetry to reduce the number of integrations by placing the vector $q_1$ along the axis $z$ and $q_2$ along the plane $x-z$ so that $q_{1,\theta}=q_{1,\phi}=q_{2,\phi} = 0$.
There are however many possible parametrizations depending on the choice of the momenta, some of which may allow further simplifications and be numerically more stable than the others. 
In particular, we notice that the scalar products of the momenta, appearing in the propagators and as arguments of the effective vertices, have a sizable impact on the evaluation speed and final precision of the computation, with the best results obtained for the integrands with fewer scalar products due to the simpler structure which limits accidental cancellations.
Moreover, for the remaining scalar products it is better to prefer those expressed in terms of $\vq_1$ and $\vq_2$ since, with our choice of coordinates, they depend on fewer angles.
Furthermore, it is important to select what functions have the scalar products as arguments, preferring the bubbles and triangles to the propagators.
Our implementation, available in the \texttt{Phi4tools} package, takes into account all these considerations, scanning different linear combinations of the internal momenta, and picking the best one.\\

In the case of $\Gamma^{(2)'}(p^2=0)$, the momentum derivative of 1PI two-point function, there is an additional step to write the integrand of a diagram because we need to take the derivative with respect to the external momentum $\vp$ before setting it to zero. We proceed as above by assigning all the  momenta, but this time also taking into account $\vp$ and making sure it appears on the least amount of effective vertices and propagators. We then evaluate the derivative of the resulting function $\widetilde{f}(\vec{p}, \vq_1,...,\vq_\ell)$ with respect to $p^2$ using the relation
\begin{equation}
    \frac{\d}{\d p^2} \left( \left. \int \prod_{i=1\dots \ell} \rd^d \vq_i \ \widetilde{f}(\vec{p},\vq_1,...,\vq_\ell) \right) \right\vert_{p=0}= \int \prod_{i=1\dots \ell} \rd^d \vq_i \ \left(\left. \frac{1}{2d} \frac{\d}{\d \vec{p}} \cdot \frac{\d}{\d \vec{p}} \ \widetilde{f}(\vec{p},\vq_1,...,\vq_\ell) \right\vert_{p=0}\right)\,.
\end{equation}
At this point one can write the integral in spherical coordinates as explained above.\\

In ref.~\cite{phi4tools:2023zen} we provide the text files containing precomputed lists of integrands for the Feynman diagrams with up to eight quartic vertices. For each listed diagram, the entry reports the Nickel index, the value of $\ell$, and the function $f$. With the \texttt{Phi4tools} package, it is possible to access the data, as well as to write down the integrands in the form $f(\vq_1,...,\vq_\ell)$ for any of the other Feynman diagrams. Moreover, 
it allows one to explicitly print the integrand as a function of the radial and angular components of the momenta in three dimensions, e.g. see figure \ref{fig:exampleWrite} in appendix.

\subsection{Performing the integration}

As the last step, we perform the integrations for the diagrams of the $\phi^4$ theory. While for simple enough diagrams one can solve the integrals analytically \cite{Nickel:1978ds,Sberveglieri:2020eko,Kudlis:2022rmt}, at higher loops we resort to numerical integration. Different numbers of effective loops correspond to different dimensions of integration, and we find that it is better to differentiate the programs used in the different cases. The case $\ell=0$ already corresponds to integrated values. For $\ell=1$ we have one-dimensional integrals that can be performed with arbitrary precision by Mathematica (although a limitation is present when numerical functions are involved) in a matter of fractions of seconds. For $\ell=2$ we have three-dimensional integrals that Mathematica can still manage, this time to achieve a precision of at least 10 significative digits we need some hours, so we run those integrals on the SISSA cluster Ulysses. For $\ell \geq 3$, corresponding to integrals with $D\geq 6$, we used the Monte Carlo VEGAS algorithm \cite{Lepage:1977sw} from the python module $\texttt{vegas}$. We run those integrals in part on the SISSA cluster Ulysses and in part on CINECA cluster Marconi, each one of them for one or two days.
The values for the computationally most difficult diagrams have been cross-checked using different parametrizations of the integrands and, in some instances, using the \verb|feyntrop| computer program \cite{Borinsky:2023jdv}.\footnote{We thank the referee for suggesting this software to us.}
The final numeric uncertainties vary considerably among the different diagrams (as in the case of ref.~\cite{Nickel:1977gu}), depending on the number of simplifications, on the type of substitutions and on the parametrization of the integrands. The largest relative uncertainties for the two- and four-point diagrams are of few parts in $10^{5}$, while the largest relative uncertainties for the derivative of the two-point diagrams reach few parts in $10^4$.
\\

The text files for each order containing the values of the Feynman diagrams can be found in ref.~\cite{phi4tools:2023zen}, each row containing the Nickel index of a diagram, its value, and the related uncertainty. The normalization that we adopted are presented in appendix \ref{app:Conventions}. With the \texttt{Phi4tools} package, it is possible to quickly have the values of any of the diagrams of the $\phi^4$ theory, e.g. see figures \ref{fig:exampleInformationD} and \ref{fig:exampleValues} in appendix.

\subsection{Numerical uncertainties and comparison with ref.~\cite{Nickel:1977gu}}

\begin{figure}[t!]
\centering
  \raisebox{0\height}{\includegraphics[width=0.47\textwidth]{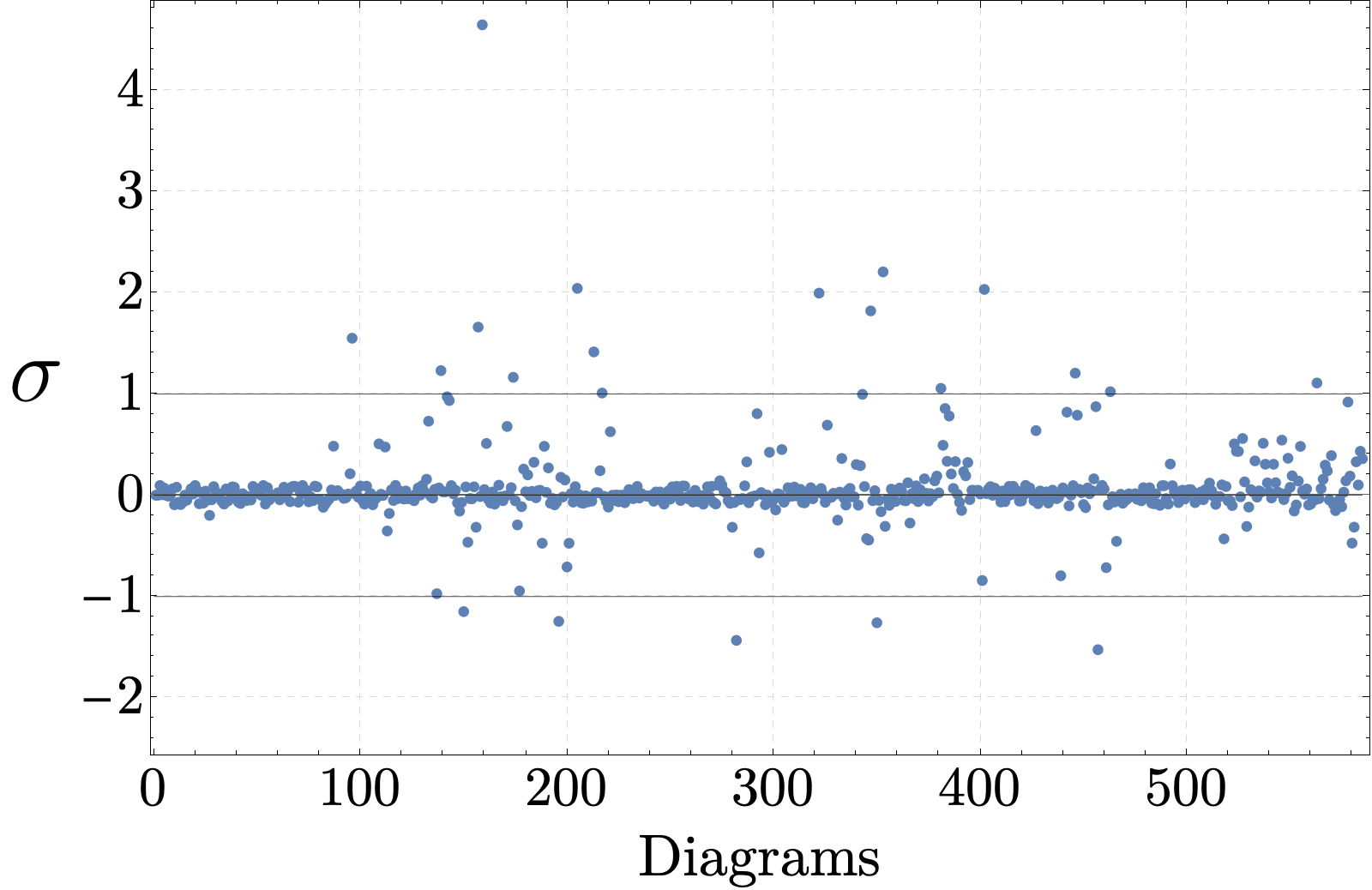}}
  \hspace*{.1cm}
  \raisebox{-0.03\height}{\includegraphics[width=0.47\textwidth]{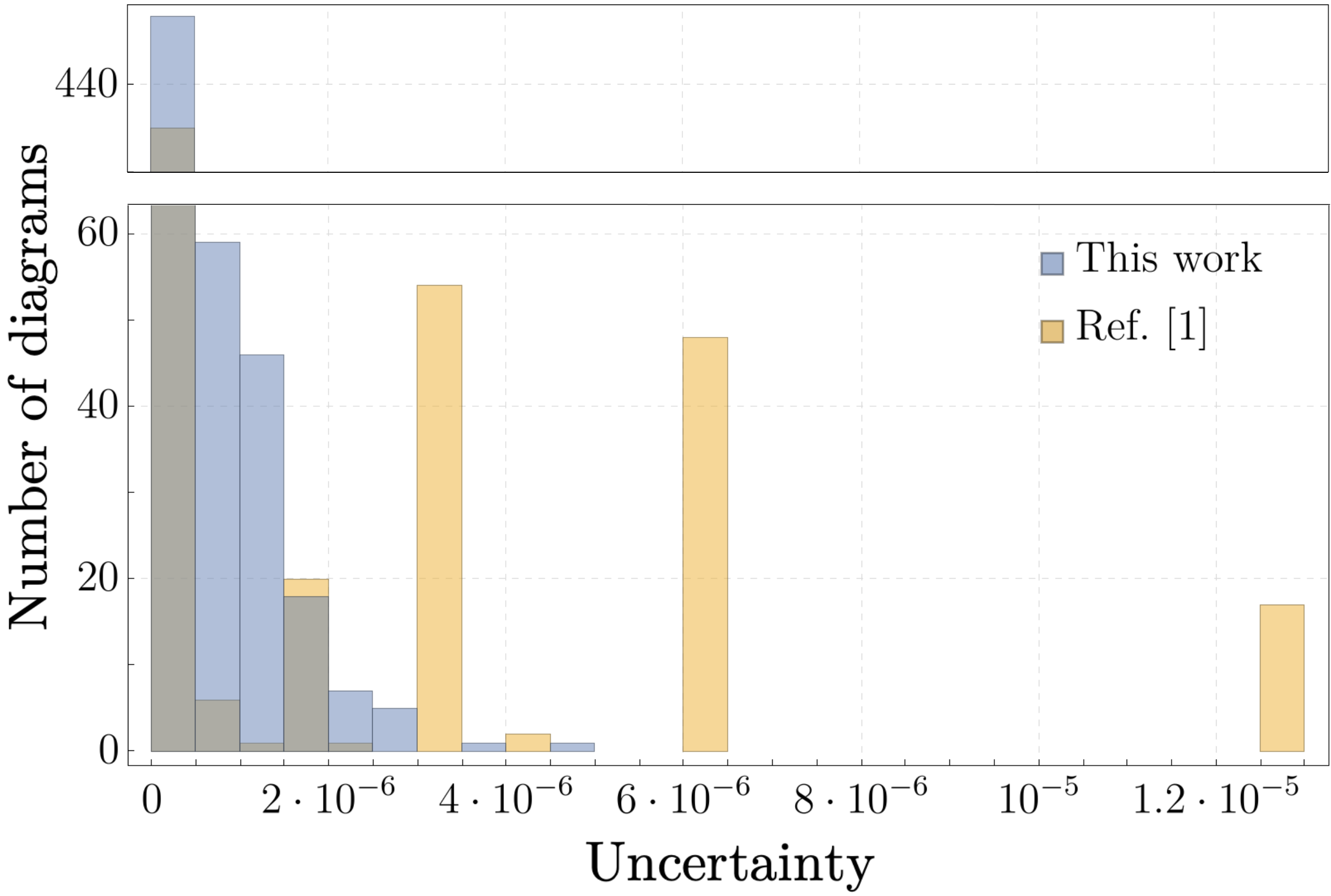}}
  \caption{Comparison between this work and ref.~\cite{Nickel:1977gu} for the six-loop diagrams of the four-point function that share the same renormalization (585 out of 660 diagrams). \emph{Left panel}: compatibility between each pair of diagrams measured in standard deviations $\sigma$. \emph{Right panel}: distribution of the uncertainties for the values normalized as explained in appendix~\ref{app:Conventions}. For the values in ref.~\cite{Nickel:1977gu} we assumed an uncertainty of $\pm 5$ on the last reported digit, but the real uncertainty might be larger (see text). One diagram of ref.~\cite{Nickel:1977gu} has uncertainty equal to $2.6 \times 10^{-5}$ and lies outside the range of the plot.}
  \label{fig:comparison}
\end{figure}

The results up to six loops can be compared with ref.~\cite{Nickel:1977gu}, providing, more than 45 years later, an independent check of the unpublished results. Before turning to the comparison, we note that in the tables of \cite{Nickel:1977gu}, the values of the diagrams reported correspond to the theory with a different renormalization counterterm, that subtracts the whole two-point subdiagrams at zero momenta instead of just the divergent tadpole and sunset diagrams.
Instead, we have chosen a minimal counterterm that only removes the divergences, and we listed the values of the two-point function explicitly. In our opinion, this simpler and more direct presentation will facilitate the reuse of the data. The result is that a small subset of the diagrams has different values in the two schemes.
In figure~\ref{fig:comparison}, we show the direct comparison for the six-loop diagrams of the four-point function that share the same renormalization. In the left panel, we report the compatibility between each pair of diagrams measured in standard deviations, while in the right panel, we report the distribution of the uncertainties. We note that the uncertainties in ref.~\cite{Nickel:1977gu} are not explicitly reported diagram by diagram, opting for a more generic statement that ``\emph{the errors are generally less than $\pm 5$ in the last digit quoted}'' and that ``\emph{in exceptional cases the error may run as high as $\pm 20$}''. The comparison in figure~\ref{fig:comparison} is performed assuming an uncertainty of $\pm 5$ in their last digit quoted.
We find a remarkable accuracy of the results of ref.~\cite{Nickel:1977gu}, with most of their values accurate even at the $\pm 1$ level in their last digit quoted and only a few exceptions. Only one value is off by $\sim 20$ in their last digit quoted, appearing above $\sigma=4$ in the left panel of figure~\ref{fig:comparison} and corresponding to the diagram with Nickel index $\text{ee}12|\text{e}34|\text{e}56|456|56|6||$. 
The distributions of the uncertainties are reported in the right panel of figure~\ref{fig:comparison}.
One could also compare the remaining subset by explicitly carrying out the subtraction of the two-point subdiagrams. An indirect check of Nickel's values up to six loops is obtained in the following section, where we find agreement between our series for the RG functions and the ones derived from his results.

\section{Series for RG functions in $d=3$}
\label{sec:resummation}

The computation of the Feynman diagrams, detailed in the previous section, provided the perturbative series for $\Gamma^{(2)}$, $\Gamma^{(2)'}$ and $\Gamma^{(4)}$ at zero external momentum for the three-dimensional $O(N)$-symmetric models and the $N$-component models with cubic-symmetric quartic interaction.
Moreover, the series for the two-point function with one mass insertion can be obtained as $\Gamma^{(1,2)}_{ij} = \partial \Gamma^{(2)}_{ij}/ \partial m^2_I$.
In this section, we use them to 
write down
the perturbative expansion for the three-dimensional RG functions in the renormalization scheme first proposed in \cite{Parisi:1980gya}.
The theory is renormalized by the zero-momentum conditions
\begin{equation}
    \begin{gathered}
        \label{eq:ParisiRS1}
        \Gamma^{(2)}_{ij}(p^2=0)=\delta_{ij} m^2 Z^{-1} \,, \quad
        \Gamma^{(2)'}_{ij}(p^2=0)=\delta_{ij}Z^{-1} \,, \quad
        \Gamma^{(4)}_{ijkl}(0)= m g_{ijkl} Z^{-2} \,,\\
        \Gamma^{(1,2)}_{ij}(p^2=0)=\delta_{ij}Z_{\phi^2}^{-1} \,,
    \end{gathered}
\end{equation}
which define the renormalization constants $Z$, $Z_{\phi^2}$ and the renormalized parameters $m^2$ and $g_{ijkl}$.
This scheme has been commonly adopted to study the critical behavior of Landau-Ginzburg-Wilson theories within the fixed-dimensional perturbation theory \cite{Pelissetto:2000ek}.
We focus on the cubic anisotropic theory, where the $N$ scalar fields interact via two quartic couplings: the first one preserving the $O(N)$ invariance, while the second breaks it to a residual discrete cubic symmetry given by the reflections and permutations of the fields.
The renormalized coupling $g_{ijkl}$ 
is specified as
\begin{equation}
    g_{ijkl} = \frac{u}{3}(\delta_{ij}\delta_{kl}+\delta_{ik}\delta_{jl}+\delta_{il}\delta_{jk})+v \ \delta_{ij}\delta_{ik}\delta_{il}\,,
\end{equation}
where $u$ and $v$ are related to the bare dimensionful parameters $u_0$ and $v_0$ of eq.~(\ref{eq:TensLCub}) through 
\begin{equation} \label{ParisiRSCubic}
    u_0=m u Z_u(u,v) Z(u,v)^{-2}\,, \qquad v_0=m v Z_v(u,v) Z(u,v)^{-2}\,.
\end{equation}
Clearly, the $O(N)$-symmetric model can be readily obtained by switching off the symmetry-breaking coupling $v$.
Starting from the series for the 1PI functions in the intermediate scheme and using the above conditions, we obtain the perturbative expansion for the RG functions $\beta_u$, $\beta_v$, $\eta$, and $\eta_{\phi^2}$ defined as
\begin{equation}
    \beta_u(u,v)=m\frac{\partial u}{\partial m} \bigg|_{u_0,v_0}\,, \qquad \beta_v(u,v)=m\frac{\partial v}{\partial m} \bigg|_{u_0,v_0}\,,
\end{equation}
\begin{equation}
    \eta(u,v)=m\frac{\partial \log Z}{\partial m} \bigg|_{u_0,v_0}\,, \qquad \eta_{\phi^2}(u,v)=m\frac{\partial \log Z_{\phi^2}}{\partial m} \bigg|_{u_0,v_0}\,,
\end{equation}
and we obtain the series for the RG function $\nu$ from the relation $\nu(u,v)=(\eta_{\phi^2}(u,v)-\eta(u,v)+2)^{-1}$.
Following the convention of refs.~\cite{Baker:1976ff, Baker:1977hp, Carmona:1999rm}, we rescale the couplings
\begin{equation}
    \widetilde u \equiv\frac{N+8}{48 \pi}u\,,\qquad\ \ \widetilde v\equiv\frac{3}{16 \pi}v\,,
\end{equation}
as well as the $\beta$-functions
\begin{equation}
    \beta_{\ut}(\ut,\vt) =\frac{N+8}{48 \pi}\beta_u(u,v)\,, \qquad\ \ \beta_{\vt}(\ut,\vt)=\frac{3}{16 \pi}\beta_v(u,v)\,.
\end{equation}
We find the following series up to the eighth order, that extend the results in the literature~\cite{Carmona:1999rm} by one order for $\beta_{\widetilde u}$ and $\beta_{\widetilde v}$ and by two orders for $\eta$ and $\eta_{\phi^2}$,\footnote{We note that for the $O(N)$ model the series were known up to order seven \cite{Guida:1998bx}.}
\begin{equation}\label{eq:betau}
\begin{split}
    \beta_{\widetilde u}(\widetilde u, \widetilde v)= - \ut +\ut^2 + \frac{3}{2} \ut\vt- \frac{4(41N+190)}{27 (N+8)^2} \ut^3-\frac{400}{81(N+8)}\ut^2\vt -\frac{92}{729}\ut\vt^2 + \\ +\ut\sum_{i+j \ge 3}  b^{(u)}_{ij}\left(\frac{9}{N+8}\right)^i\ut^i\vt^i\,,
\end{split}
\end{equation}
\begin{equation}\label{eq:betav}
\begin{split}
    \beta_{\widetilde v}(\widetilde u, \widetilde v)= - \vt +\vt^2 + \frac{12}{N+8} \ut\vt- \frac{308}{729} \ut^3-\frac{832}{81(N+8)}\ut^2\vt-\frac{4(23N+370)}{27(N+8)^2}\ut\vt^2 +\\
    + \vt\sum_{i+j \ge 3}  b^{(v)}_{ij}\left(\frac{9}{N+8}\right)^i\ut^i\vt^i\,,
\end{split}
\end{equation}
\begin{equation}\label{eq:eta}
    \eta(\widetilde u, \widetilde v)=\frac{8(N+2)}{27(N+8)^2}\ut^2 + \frac{16}{8(N+8)} \ut\vt+ \frac{8}{729} \vt^3 +\sum_{i+j \ge 3}  e^{(\phi)}_{ij}\left(\frac{9}{N+8}\right)^i\ut^i\vt^i\,,
\end{equation}
\begin{equation}\label{eq:eta2}
    \eta_{\phi^2}(\widetilde u, \widetilde v)= -\frac{N+2}{N+8} \ut -\frac{1}{3}\vt+\frac{2(N+2)}{(N+8)^2}\ut^2 + \frac{4}{3(N+8)} \ut\vt+\frac{2}{27}\vt^2+\sum_{i+j \ge 3}  e^{(\phi^2)}_{ij}\left(\frac{9}{N+8}\right)^i\ut^i\vt^i\,,
\end{equation}
where the coefficients $b^{(u)}_{ij}$, $b^{(v)}_{ij}$ ($3\le i+j \le 7$) and $e^{(\phi)}_{ij}$, $e^{(\phi^2)}_{ij}$ ($3\le i+j \le 8$) are reported in the tables in appendix \ref{app:CoeffCub}. 
We perform various consistency checks on our series, including those reported in ref.~\cite{Carmona:1999rm}, that involve non-trivial identities among the RG functions for $N=1$ and $N=2$.
We also find agreement between our coefficients and those in ref.~\cite{Carmona:1999rm}.
Setting $\vt=0$ we get the series for the $O(N)$-symmetric models, explicitly reported in appendix \ref{app:coeff_O(N)} as the series for $\beta_{\ut}(\ut,0)$, $\eta(\ut,0)$, and $\nu(\ut,0)$.
We have carefully checked these series, comparing the coefficients for $\eta$ and $\eta_{\phi^2}$ up to order $\gtni^7$ for $N=0,1,2$, and $3$ with those computed by Murray and Nickel and appearing in the appendix of ref.~\cite{Guida:1998bx}. We also compared the coefficients of $\beta$ with those in ref.~\cite{Baker:1977hp}, finding  agreement in all cases.\\
After resummation, the longer perturbative series are expected to deliver more precise results for the fixed-dimensional estimates of the critical exponents, which could then be compared with the estimates from $\epsilon$-expansion \cite{Guida:1998bx,Carmona:1999rm,Kompaniets:2017yct,Adzhemyan:2019gvv}, Monte Carlo simulations \cite{Pelissetto:2007tx, Hasenbusch_2010, Campostrini:2006ms, Campostrini:2002ky} and Conformal Bootstrap \cite{Kos:2016ysd, Chester:2019ifh, Chester:2020iyt}. We postpone such study to a future work.

\section{Conclusions}
\label{sec:conclusions}

With this work, we have provided a comprehensive 
collection of results
for fixed-dimensional perturbative computations for theories described by Landau-Ginzburg-Wilson Hamiltonians.
In particular, we have illustrated the steps of our computations and shared, for each of them, the results and the tools developed.
They range from the construction and labeling of the Feynman diagrams, the computation of the symmetry factors for $O(N)$ and cubic anisotropic theories, the compilation of the integrands in momentum space, and finally to their numerical evaluation in the specific case of three-dimensional quartic theories. 
The fundamental operation that enabled us to push the computation to high orders was the identification and replacement of specific subdiagrams with their analytical expression. This operation, done systematically at the diagrammatic level, 
allowed us to reduce the number of loops of the Feynman diagrams and, in turn, the dimension of the integration space, leading to the evaluation of all the Feynman diagrams with up to eight quartic vertices for the zero, two, and four-point functions in the three-dimensional theory at zero external momentum.
All the results are available in text files in ref.~\cite{phi4tools:2023zen}, and can be accessed through \texttt{Phi4tools}, 
a handy and easy-to-navigate Mathematica paclet with a comprehensive documentation that is aimed at anyone who needs to consult a single source for information on this topic. 
Our results constitute an independent check for the unpublished work of ref.~\cite{Nickel:1977gu}.
We then presented some applications of the computed diagrams, obtaining the perturbative series for RG functions for the $O(N)$ and the cubic anisotropic model to an unprecedented high order.
The resummation of the series, which will be carried out in a future work, will give the state-of-the-art fixed-dimensional estimates of the critical exponents, that could be benchmarked against other methods, similarly to what was done in ref.~\cite{Bonanno:2022ztf}.
Moreover, the perturbative series for other models can be easily implemented using the package by specifying the appropriate symmetry factors for the diagrams. It is also possible to extend the computations in various directions, with a number of diagrams already implemented but not computed.
The diagrams with nine quartic vertices are already present, with multiplicities and integrands with the substitutions of the effective vertices in place. 
The same is valid for the Feynman diagrams with cubic vertices with up to eight total vertices, cubic plus quartic.

\acknowledgments

We are grateful to Marco Serone for his encouragement on the project and for many helpful and insightful discussions throughout the course of this work.
We also thank Nicola Dondi, Stefano Giorgini, and Saman Soltani for useful discussions and comments on the draft.
G.Sp.~acknowledges support from the Provincia Autonoma di Trento. 
The work of G.Sb.~is supported by the Swiss National Science Foundation under grants number 205607, within the scope of the NCCR SwissMAP, and 200021~192137. We acknowledge the HPC Collaboration Agreement between SISSA and CINECA for granting access to Marconi100. Part of the numerical computations were performed on the SISSA cluster Ulysses.

\appendix
\section{Conventions and normalizations}
\label{app:Conventions}

This appendix 
is devoted to spelling out the conventions used in this work, in particular for the normalization for the symmetry and multiplicity factors, the integrands, and the values of the diagrams in three dimensions. 
These normalizations can also be found in the documentation of the \texttt{Phi4tools} paclet, and the computation of the series for the $\beta$-function can be found, step by step, in the Tech Notes.\\

Let's consider a generic Feynman diagram $\mathcal{G}$ of an $n$-point function with $v_3$ cubic vertices and $v_4$ quartic ones, $\mathcal{G}$ is identified by its Nickel index $\mathfrak{N}\left(\mathcal{G}\right)$. Its \textit{multiplicity factor} $M_\mathcal{G}$ is equal to the number of Wick contractions leading to this same diagram. We define \textit{weight factor} $W_\mathcal{G}$ as
\begin{equation}
      W_\mathcal{G}=\frac{M_\mathcal{G}}{v_3!(3!)^{v_3}v_4!(4!)^{v_4}}\,.
\end{equation}
The integrand for $\mathcal{G}$ that can be obtained from the paclet is the product between the factors coming from the propagators and the effective vertices and $W_\mathcal{G}$ (thus without the usual integral measure $(2\pi)^{-d}$ in $d$ dimensions for every loop). The weight factor lacks the contribution due to the number of different channels for a given diagram. We call it $M_{\text{ch}}(n)$ since it depends only on the number of external legs. For the zero and two-point functions we have $M_{\text{ch}}(0)=M_{\text{ch}}(2)=1$, while for the four-point function we have $M_{\text{ch}}(4)=3$. Hence, for example, the weight factor for the one-loop bubble diagram in the four-point function with two quartic vertices, labeled by the Nickel index $\text{ee}11|\text{ee}|$, is given by $W_{\includegraphics[scale=0.04]{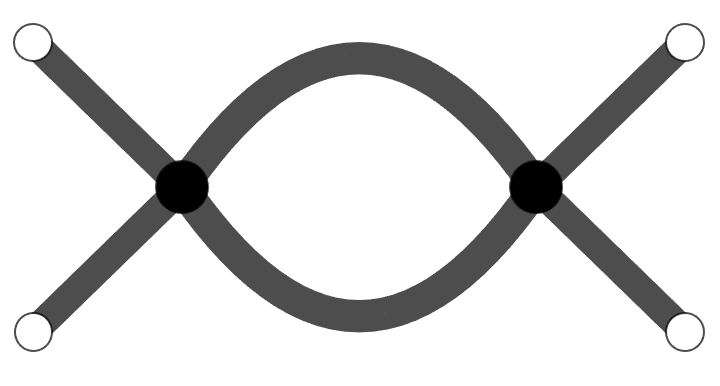}}=\frac{1}{2}$.\\

As explained in the main text, the paclet also provides the \textit{symmetry factors}\footnote{Notice that in literature the term symmetry factor is sometimes used to indicate the inverse of what we call the weight factor $W_\mathcal{G}$. In our conventions, the symmetry factor indicates the factor coming from the tensorial structure of the $\phi^4$ interaction.} coming from the tensorial structure of $\lambda_{ijkl}$ for the $O(N)$ model and the anisotropic cubic model, see eq.~\eqref{eq:TensLCub}.\footnote{We refer the reader to chapter 6 of ref.~\cite{Kleinert:2001critical} for a more detailed treatment.
} We indicate with $S_{\mathcal{G}}(N,X)$ the cubic symmetry factor for the diagram $\mathcal{G}$ where we defined $X=v_0/u_0$. Thus, $S_{\mathcal{G}}(N,0)$ will be the one for the $O(N)$ model.
If $\mathcal{G}$ is a diagram contributing to the two-point function ($n=2$), 
the symmetry factors $S_{\mathcal{G}}(N,X)$ correspond to the coefficient of the tensorial structure $\delta_{ij}$.
Instead, if $\mathcal{G}$ is a diagram contributing to the four-point function ($n=4$), 
the symmetry factors $S_{\mathcal{G}}(N,X)$ correspond to the list of the coefficients that multiply the four possible channels $(\delta_{ij}\delta_{kl},\delta_{ik}\delta_{jl},\delta_{il}\delta_{jk},\delta_{ij}\delta_{ik}\delta_{il})$;
upon symmetrization, 
we obtain $S^{\text{sym}}_{\mathcal{G}}(N,X)$ given by the list of the two coefficients of the tensorial structures $(\left(\delta_{ij}\delta_{kl}+\delta_{ik}\delta_{jl}+\delta_{il}\delta_{jk}\right)/3,\delta_{ij}\delta_{ik}\delta_{il})$, where the first coefficient is given simply by the sum of the first three of $S_{\mathcal{G}}(N,X)$. The operation of symmetrization instead acts as the identity on the diagrams with $n=0$ and $n=2$, thus, in that case, $S^{\text{sym}}_{\mathcal{G}}(N,X)$ coincides with $S_{\mathcal{G}}(N,X)$. 
The symmetry factors are normalized so that for the Ising model they are equal to 1, i.e., for every $\mathcal{G}$, we have 
\begin{align}
    & S^{\text{sym}}_{\mathcal{G}}(1,0) =1  \hspace{1.85cm} \text{for} \qquad  n=0,2\,, \\
    & S^{\text{sym}}_{\mathcal{G}}(1,0) = (1,0) \hspace{1.2cm} \text{for} \qquad n=4\,.
\end{align}
As an example, we report below the symmetry factor for the diagram with the label $\text{ee}11|\text{ee}|$, together with its symmetrized version,
\begin{align}
    S_{\includegraphics[scale=0.04]{figs/im_bubble4.png}}(N,X) &= 
      \left(\frac{N+4+6X}{9},\frac{2}{9},\frac{2}{9},\frac{4X}{3}+X^2\right)
    \,, \\
    S^{\text{sym}}_{\includegraphics[scale=0.04]{figs/im_bubble4.png}}(N,X) &= 
      \left(\frac{N+8+6X}{9},\frac{4X}{3}+X^2\right)\,.
\end{align}

Let's call $\mathcal{V}_{\mathcal{G}}$ the value of the Feynman diagram $\mathcal{G}$ in three dimensions in the renormalization scheme of eq.~\eqref{eq:dm2I}. The quantity $\mathcal{V}_{\mathcal{G}}$ is calculated by integrating the integrand for $\mathcal{G}$, normalized as explained earlier, and multiplying the results by $(16\pi)^l$, where $l$ is the number of loops of $\mathcal{G}$. In this way, the values for most of the diagrams for the four-point function are of order one.\footnote{Our normalization is similar to the one used in ref.~\cite{Nickel:1977gu}. For the diagrams not affected by the difference in normalization scheme, we have that $\mathcal{V}_{\mathcal{G}}$ is equal to their value times what they call $Z_1$ symmetry factor.} For our usual example with label $\text{ee}11|\text{ee}|$, the value is given by $\mathcal{V}_{\includegraphics[scale=0.04]{figs/im_bubble4.png}}=1$.\\

Finally, let us show, as an example, how to write the perturbative series for 1PI $n$-point functions at zero external momenta in the $O(N)$ model. We have 
\begin{equation}\label{eq:Gamma_n_series}
    \Gamma^{(n)} = m_I^{(3-\frac{n}{2})}\sum_k \sum_{\mathcal{G}^{(k)}} (-1)^{(k-1)} M_{\text{ch}}(n) (16\pi)^{-(k+1-\frac{n}{2})} \ S^{\text{sym}}_{\mathcal{G}^{(k)}}(N,0) \mathcal{V}_{\mathcal{G}^{(k)}} \left(\frac{u_0}{m_I}\right)^k\,,
\end{equation}
where $\mathcal{G}^{(k)}$ indicates a Feynman diagram with $v_4=k$.
We composed the symmetry factors\footnote{For $n=4$, the symmetry factors $S^{\text{sym}}_{\mathcal{G}^{(k)}}$ are actually a list of two values, one corresponding to the $O(N)$-symmetric tensorial structure and the other corresponding to the $O(N)$-breaking tensorial structure. With a slight abuse of notation, in the $O(N)$ example of eq.~\eqref{eq:Gamma_n_series}, we assume we just take the first of the two coefficients, the other being zero. In the more general case, one should instead take the sum of the two, each multiplied by the corresponding tensorial structure.} and values, summed over all the diagrams at a given order $p$, and fixed the normalization: the factors $16\pi$ are there to balance the normalization of $\mathcal{V}_{\mathcal{G}^{(k)}}$, and we have the factor $M_{\text{ch}}(n)$ accounting for the number of channels.
The series for the more general case of the cubic anisotropic model can be obtained in a similar fashion.

\section{Effective vertices in $d=3$}
\label{app:eff_vertices}

In this appendix, we report functions and analytic terms corresponding to the effective vertices presented in section \ref{sec:substitutions} in three dimensions. In order to make the expressions lighter, we put $m_I=1$ in this appendix. The masses can be easily reinserted using dimensional analysis. We refer the reader to ref.~\cite{Nickel:1978ds} for the formulas for one-loop subdiagrams with propagators with different masses. 
We have for the \textit{bubble} diagram 
\begin{equation}
    \mathcal{B}(p)=\frac{\arctan(p)}{4\pi p}\,.
\end{equation}
For the \textit{sunset} in our renormalization scheme, i.e. $\mathcal{S}(p=0)=0$, we have
\begin{equation}
    \mathcal{S}(p)=-\frac{1}{32\pi^2}\left(\log\left(\frac{p^2}{9}+1\right)+\frac{6\arctan(\frac{p}{3})}{p}-2\right)\,.
\end{equation}
The \textit{triangle} diagram has three external legs, so it depends on $(\vp_1,\vp_2,\vp_3)$. However, since $\vp_1+\vp_2+\vp_3=0$, it will be a function of two vectors or three scalars.\footnote{In our notation, $\vec{p}$ indicates the three-dimensional vector and $p\equiv |\vec{p}|$ its magnitude.}
Hence, we can write
\begin{equation}
    \mathcal{T}(p_1,p_2,p_3)=\frac{\arctan\left(\frac{\sqrt{\mathcal{D}_3(p_1,p_2,p_3)}}{\frac{p_1^2+p_2^2+p_3^2}{2}+4}\right)}{8 \pi 
   \sqrt{\mathcal{D}_3(p_1,p_2,p_3)}}\,,
\end{equation}
with
\begin{equation}
    \mathcal{D}_3(p_1,p_2,p_3)=\frac{1}{4} \left(p_1^2 p_2^2 p_3^2+(p_1+p_2-p_3)(p_1-p_2+p_3)(-p_1+p_2+p_3)(p_1+p_2+p_3)\right)\,.
\end{equation}
We can also write $\mathcal{D}_3$ as
\begin{equation}
 \mathcal{D}_3(p_1,p_2,p_3)=\det D^{(3)}(p_1,p_2,p_3)\, \qquad \text{with} \qquad D^{(3)}(p_1,p_2,p_3)=
\begin{pmatrix}
1 & 1+\frac{p_2^2}{2} & 1+\frac{p_1^2}{2}\\
1+\frac{p_2^2}{2} & 1 & 1+\frac{p_3^2}{2}\\
1+\frac{p_1^2}{2} & 1+\frac{p_3^2}{2} & 1
\end{pmatrix}.
\end{equation}
The \textit{square} diagram has four external legs, so it depends on $(\vp_1,\vp_2,\vp_3,\vp_4)$. However, since $\vp_1+\vp_2+\vp_3+\vp_4=0$, it will be a function of three vectors or six scalars. Let us introduce
\begin{equation}
D^{(4)}(p_1,p_2,p_3,p_4,|\vp_1+\vp_2|,|\vp_2+\vp_3|)=
\begin{pmatrix}
1 & 1+\frac{p_1^2}{2} & 1+\frac{\left(\vp_1+\vp_2\right)^2}{2} & 1+\frac{p_4^2}{2}\\
1+\frac{p_1^2}{2} & 1 & 1+\frac{p_2^2}{2} & 1+\frac{\left(\vp_2+\vp_3\right)^2}{2}\\
1+\frac{\left(\vp_1+\vp_2\right)^2}{2} & 1+\frac{p_2^2}{2} & 1 & 1+\frac{p_3^2}{2}\\
1+\frac{p_4^2}{2} & 1+\frac{\left(\vp_2+\vp_3\right)^2}{2} & 1+\frac{p_3^2}{2} & 1
\end{pmatrix} 
\end{equation}
and 
\begin{equation}
\mathcal{D}_4=\det D^{(4)}\left(p_1,p_2,p_3,p_4,|\vp_1+\vp_2|,|\vp_2+\vp_3|\right)\,.
\end{equation}
Let's now define the principal minors $\mathcal{D}_{3,i}$ obtained by
eliminating the $i^{\text{th}}$ row and column from $D^{(4)}$ and $\mathcal{F}_{4,i}$ as the determinant of the $4 \times 4$ matrix obtained from $D^{(4)}$ and replacing the elements of the $i^{\text{th}}$ column by ones.
Concretely, with our parametrization, we have 
\begin{align}
    \mathcal{D}_{3,1} & \equiv \mathcal{D}_3(p_2,p_3,|\vp_2+\vp_3|)\,, \quad 
    \mathcal{D}_{3,2} \equiv \mathcal{D}_3(|\vp_1+\vp_2|,p_3,p_4)\,, \\
    \mathcal{D}_{3,3} & \equiv \mathcal{D}_3(p_1,|\vp_2+\vp_3|,p_4)\,, \quad
    \mathcal{D}_{3,4} \equiv \mathcal{D}_3(p_1,p_2,|\vp_1+\vp_2|)\,.
\end{align}
Let's use the same labeling for the triangle functions, i.e.
\begin{align}
    \mathcal{T}_{1} & \equiv \mathcal{T}(p_2,p_3,|\vp_2+\vp_3|)\,, \quad 
    \mathcal{T}_{2} \equiv \mathcal{T}(|\vp_1+\vp_2|,p_3,p_4)\,, \\
    \mathcal{T}_{3} & \equiv \mathcal{T}(p_1,|\vp_2+\vp_3|,p_4)\,, \quad
    \mathcal{T}_{4} \equiv \mathcal{T}(p_1,p_2,|\vp_1+\vp_2|)\,.
\end{align}
Finally, we have
\begin{equation}
   \mathcal{Q}(p_1,p_2,p_3, p_4,|\vp_1+\vp_2|,|\vp_2+\vp_3|) = \frac{1}{2\mathcal{D}_{4}}\sum_{i=1}^4 \mathcal{F}_{4,i}  \mathcal{T}_i\,.
\end{equation}
For the \textit{effective tadpoles}, we have
\begin{equation}
    t_{\mathcal{S}}=-\frac{\log\left(\frac{4}{3}\right)}{128 \pi^3}\,, \qquad\qquad\qquad t_{\mathcal{TB}}= \frac{1}{12288 \pi^2}\,.
\end{equation}
Lastly,
\begin{equation}
    \mathcal{T}_{bA}(p)=\frac{\arctan\left(\frac{p}{3}\right)}{32\pi p}\,.
\end{equation}

\section{The \texttt{Phi4tools} paclet}
\label{app:paclet}

\texttt{Phi4tools} is a Mathematica paclet that extends the Wolfram System with new functionality and includes Wolfram Language functions, documentation, and data files.
It is published in the Wolfram Paclet Repository\footnote{\url{https://resources.wolframcloud.com/PacletRepository/resources/GSberveglieri/Phi4tools}.} 
and it can be installed in Mathematica by evaluating the command
\begin{verbatim}
    PacletInstall["GSberveglieri/Phi4tools"]
\end{verbatim}
and then it can be loaded with the command
\begin{verbatim}
    Needs["GSberveglieri`Phi4tools`"]
\end{verbatim}
After doing so, the new symbols and functions are integrated into the system together with the related documentation, consisting of a guide page with the overview of the package, two tutorials, and the documentation pages associated with each new symbol that can be accessed from any notebook using the command \texttt{Information} (or simply ``\texttt{??}"). The documentation can also be consulted on the web from the Wolfram Paclet Repository.
We refer the reader to the guide page for a complete and comprehensive description of all functionalities, and we report below just some basic examples. For more advanced examples with the usage of multiple functions, we invite the reader to consult the Tech Notes of the paclet.

\begin{figure}[h!]
\centering
  \includegraphics[width=1.05\textwidth]{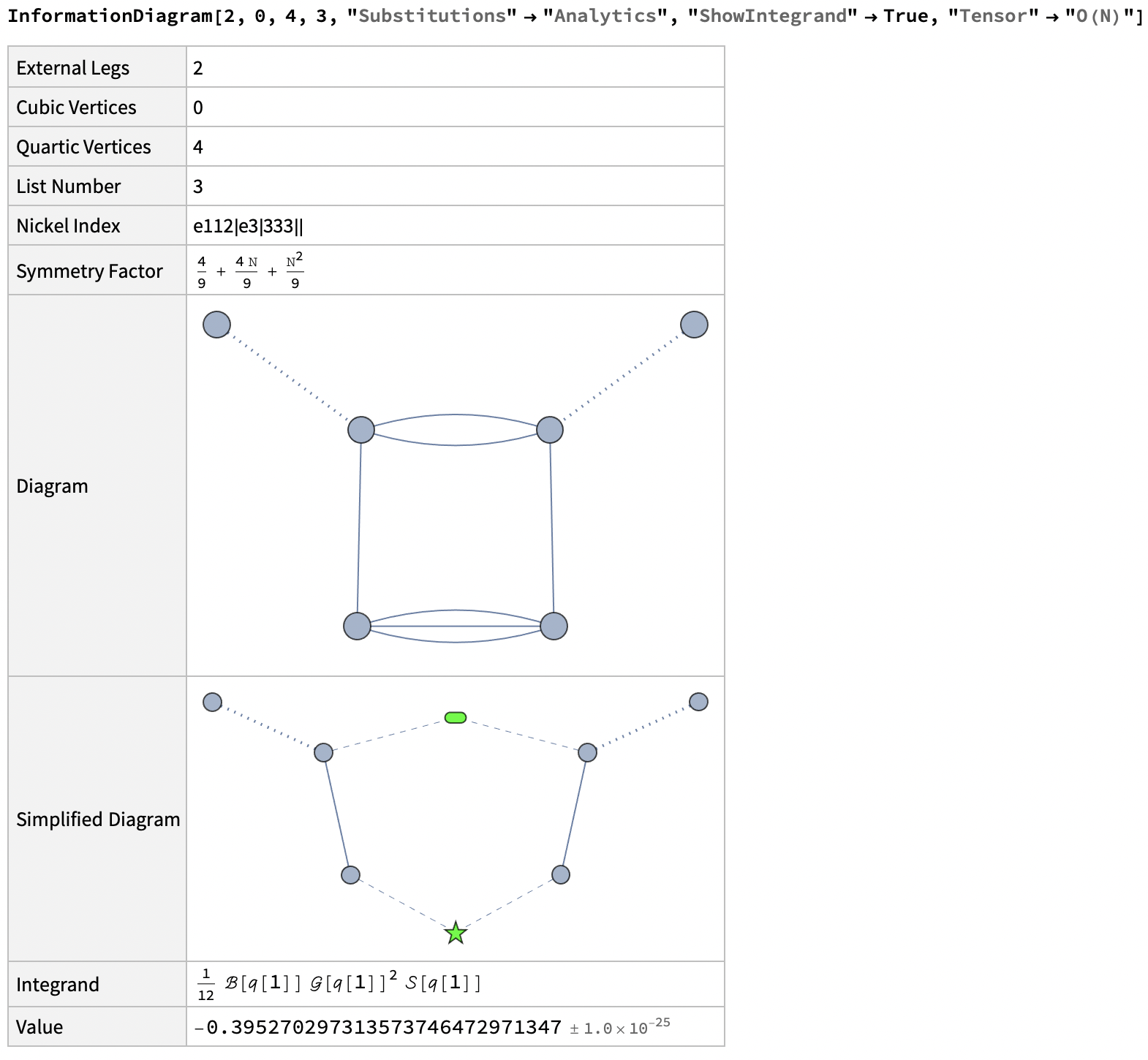}
  \caption{Example of input and output of the function \texttt{InformationDiagram} in Mathematica. The chosen diagram has Nickel index $\text{e}112|\text{e}3|333||$. It is the third of those with 4 quartic vertices of $\Gamma^{(2)}$, using the order explained in section \ref{sec:labeling} and excluding the diagrams with tadpoles. The options are chosen so that the simplified diagrams, the relative integrand, and the symmetry factor are printed out. The value refers to the three-dimensional theory. For more detailed information on symbols, options, and normalization, we refer the reader to the documentation of the paclet.}
  \label{fig:exampleInformationD}
\end{figure}

\begin{figure}[h!]
\centering
  \includegraphics[scale=0.6]{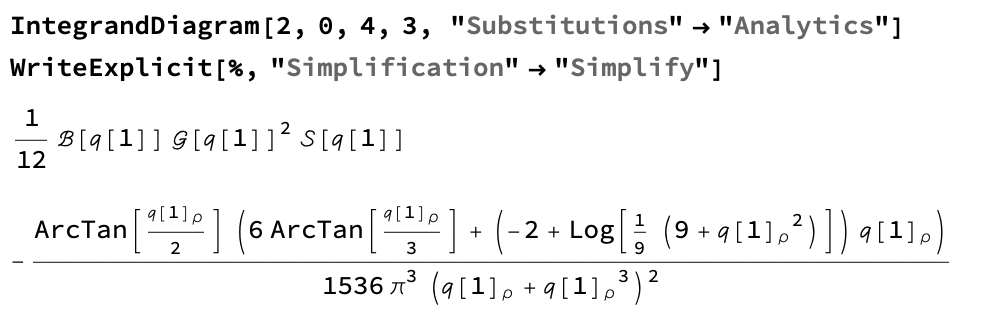}
   \caption{Example of input and output of the functions \texttt{IntegrandDiagram} and \texttt{WriteExplicit} in Mathematica.}
  \label{fig:exampleWrite}
\end{figure}

The function \texttt{InformationDiagram}$[n,v3,v4,d]$ gives details about the $d$-th diagram of the $n$-point function $\Gamma^{(n)}$ with $v3$ cubic vertices and $v4$ quartic vertices at zero external momentum. It works for zero, two, and four-point functions and provides, when computed, the value of the diagram in three dimensions. \texttt{InformationDiagram} can take as input alternatively the Nickel index or the graph associated with the desired diagram. In figure \ref{fig:exampleInformationD}, we report an example.
With \texttt{WriteExplicit}$[integrand]$, the integrand is written in the form ready to be integrated in three dimensions. In figure \ref{fig:exampleWrite}, we report an example using the same diagram of figure \ref{fig:exampleInformationD}.
\begin{figure}[h!]
\centering
  \includegraphics[width=1.04\textwidth]{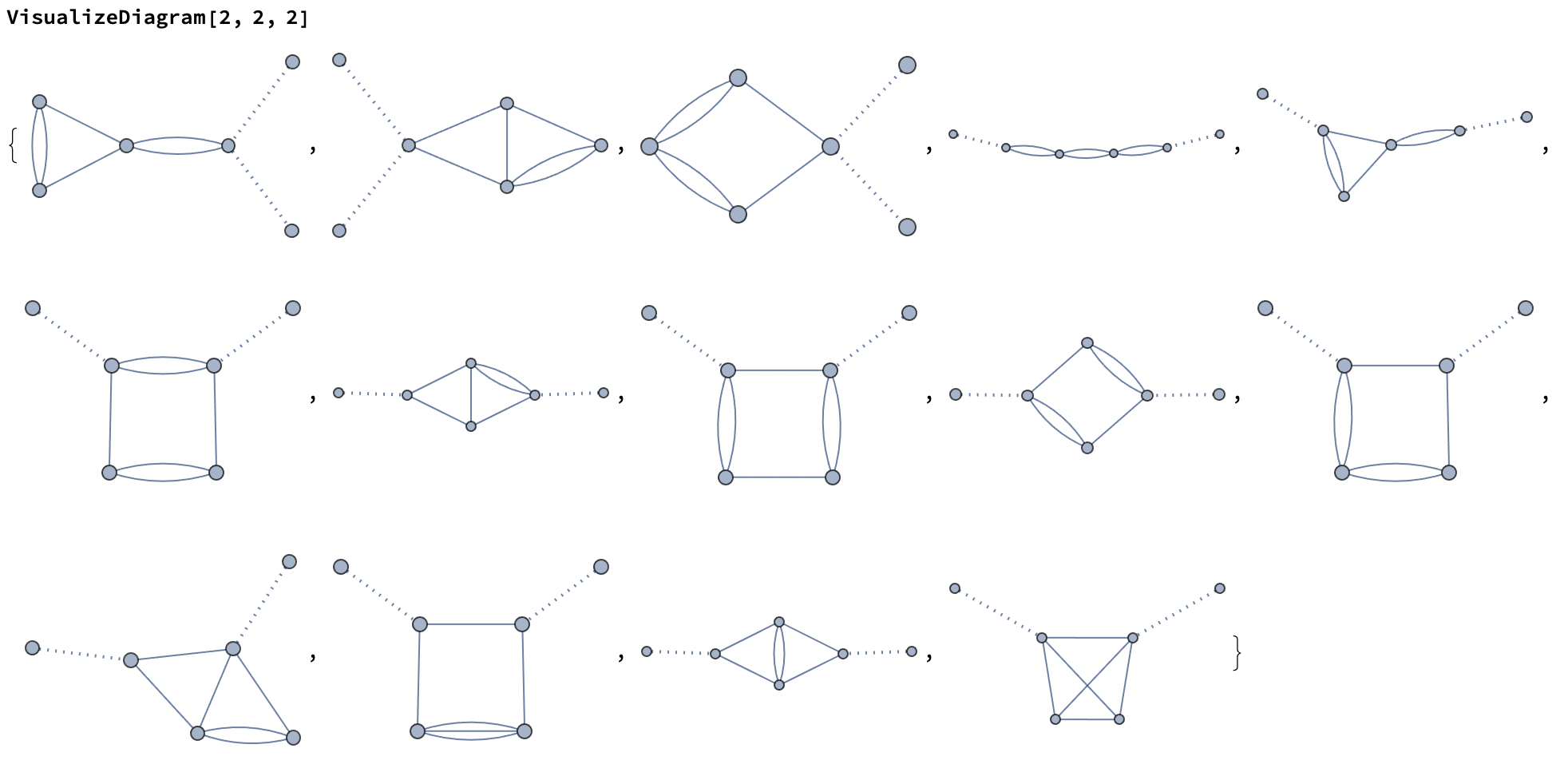}
     \caption{Example of input and output of the functions \texttt{VisualizeDiagram} for all the diagrams with 2 cubic vertices and 2 quartic ones of $\Gamma^{(2)}$.}
  \label{fig:exampleDiagrams}
\end{figure}
The function \texttt{VisualizeDiagram}$[n,v3,v4]$ draws all Feynman diagrams of the $n$-point function $\Gamma^{(n)}$ with $v3$ cubic vertices and $v4$ quartic vertices. In figure \ref{fig:exampleDiagrams}, we report an example.
\begin{figure}[h!]
\centering
  \includegraphics[width=1.08\textwidth]{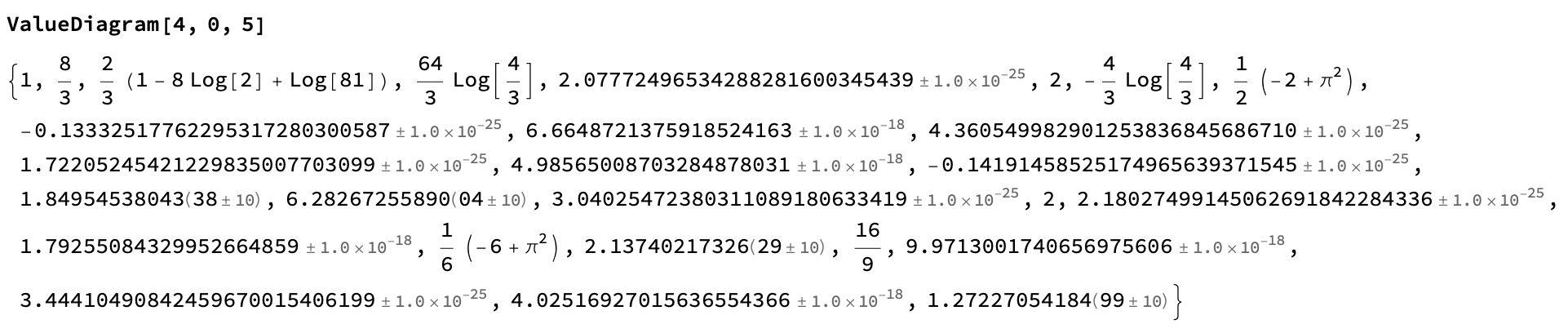}
     \caption{Example of input and output of the functions \texttt{ValueDiagram} for all the diagrams with 0 cubic vertices and 5 quartic vertices of $\Gamma^{(4)}$. The values refer to the three-dimensional theory. For more detailed information on the normalization, we refer the reader to the documentation of the function in the paclet.}
  \label{fig:exampleValues}
\end{figure}
The function \texttt{ValueDiagram}$[n,v3,v4]$ gives the list of three-dimensional integrated values of the Feynman diagram for the $n$-point function $\Gamma^{(n)}$ with $v3$ cubic vertices and $v4$ quartic vertices. In figure \ref{fig:exampleValues}, we report an example.

\FloatBarrier

\section{Series coefficients in the $O(N)$ model}
\label{app:coeff_O(N)}
In this appendix, we report the perturbative series of the RG functions $\beta$, $\eta$, and $\nu$ for the $O(N)$ model at generic $N$, obtained setting $\vt = 0$ as explained in the text. The numerical coefficients appearing without error have been computed to a higher accuracy and have been truncated at $10^{-15}$.

\begin{footnotesize}
\begin{align*}
    \beta_{\ut}(\gtni,0)
      =\: & - \gtni + \gtni^2 - \frac{\gtni^3}{(N+8)^2}  \frac{4 (41 N+190)}{27}
      + \frac{\gtni^4}{(N+8)^3} \bigg[1.348942760866478 \ N^2 \\ &\hspace*{1cm}
      +54.940377049302200 \ N
      +199.640417221105907
        \bigg]
      \\
      & - \frac{\gtni^5}{(N+8)^4} \bigg[
        -0.155645907585201 \ N^3
        +35.82020347182(7) N^2
        +602.5212285602(6) N\\ &\hspace*{1cm}
        +1832.2067281779(14)
        \bigg]
      \\
      & + \frac{\gtni^6}{(N+8)^5} \bigg[
      0.051236212811530\ N^4
        +3.237874(11) N^3
        +668.55456(24) N^2 \\ &\hspace*{1cm}
        +7819.5673(20) N
        +20770.183(5)
        \bigg]
      \\
      &-\frac{\gtni^7}{(N+8)^6} \bigg[
        -0.023424226049759 \ N^5
        -1.07182(8) N^4
        +265.8411(20) N^3\\ &\hspace*{1cm}
        +12669.295(24) N^2
        +114181.79(13) N
        +271300.61(28)
        \bigg]
      \\
      &+ \frac{\gtni^8}{(N+8)^7}\bigg[
        0.012640642324067 \  N^6
        +0.5433(5) N^5
        -14.386(16) N^4
        +8828.74(25) N^3\\ &\hspace*{1cm}
        +(246972.5 \pm 2.0) N^2
        +(1840997 \pm 8) N
        +(3981620 \pm 14)
        \bigg]\,.
\end{align*}
\\

\begin{align*}
    \eta(\gtni,0)
      =\: &   \frac{\gtni^2}{(N+8)^2}  \frac{8 (N+2)}{27}
        + \frac{\gtni^3}{(N+8)^3} \bigg[0.0246840009259343 \  (N^2+10 N + 16) \bigg]
      \\
      &+ \frac{\gtni^4}{(N+8)^4} \bigg[-0.004298563333341 \ N^3
	+0.667985910868(20) N^2
	+4.60922100685(3) N \\ &\hspace*{1cm}
	+6.51210986356(18)
        \bigg]
        \\
      &- \frac{\gtni^5}{(N+8)^5} \bigg[0.006550923035200 \ N^4
	-0.13245107140(8) N^3
	+1.891116(10) N^2 \\ &\hspace*{1cm}
	+15.18794(6) N
	+21.64700(9)
        \bigg]
         \\
      &+ \frac{\gtni^6}{(N+8)^6} \bigg[-0.005548920737435\ N^5
    -0.02039935040(31) N^4
   +3.05407(7) N^3 \\ &\hspace*{1cm}
   +64.0777(8) N^2
   +300.7218(34) N
   +369.714(4)
        \bigg]
         \\
      &- \frac{\gtni^7}{(N+8)^7} \bigg[0.004390810855773 \ N^6
   +0.0612032025(13) N^5
   -1.2705(4) N^4
   +35.311(7) N^3 \\ &\hspace*{1cm}
   +751.79(5) N^2
   +3345.53(18) N
   +3988.40(21)
        \bigg]
          \\
      &+ \frac{\gtni^8}{(N+8)^8} \bigg[-0.003473417276666 \ N^7
   -0.070431737(5)  N^6
   +0.151(3) N^5
   +11.6(1) N^4  \\ &\hspace*{1cm}
   +(1111\pm 1.5) N^3
   +(13674\pm 9) N^2
   +(52140\pm 27) N
   +(58297\pm 30)
        \bigg]\,.
\end{align*}\\

\begin{align*}
     \nu(\gtni,0)
      =\: &   \frac{1}{2} +   \frac{\gtni}{(N+8)}  \frac{(N+2)}{4} + \frac{\gtni^2}{(N+8)^2}  \frac{(27 N^2+16 N-76)}{216}
       + \frac{\gtni^3}{(N+8)^3} \bigg[\frac{N^3}{16}
       +0.357987483753229 \ N^2 \\ &\hspace*{0.5cm}
       + 2.20259658610878 \ N
       + 3.473243237204659 \bigg]
      \\
      &- \frac{\gtni^4}{(N+8)^4} \bigg[-\frac{N^4}{32}
	-0.399474877149036 \ N^3
	+0.609039560891(5) N^2
	+14.977065564467(32) N\\ &\hspace*{1cm}
	+24.82217386818(4)
        \bigg]
        \\
      &+ \frac{\gtni^5}{(N+8)^5} \bigg[\frac{N^5}{64}
	+0.333425900294786 \ N^4
	+0.7393030(9) N^3
	+27.122009(18) N^2 \\ &\hspace*{1cm}
	+184.70641(11)N
	+262.00439(15)
        \bigg]
         \\
      &- \frac{\gtni^6}{(N+8)^6} \bigg[-\frac{N^6}{128}
         -0.245952216148592 \ N^5
         -1.875537(5) N^4
         +0.87861(19) N^3 \\ &\hspace*{1cm}
         +417.7907(19) N^2
         +2385.941(7) N
         +3130.387(8)
        \bigg]
         \\
      &+ \frac{\gtni^7}{(N+8)^7} \bigg[\frac{N^7}{256}
          +0.170195139781328 \ N^6
         +2.298000(25) N^5
         +0.1341(20) N^4
         +328.511(29) N^3  \\ &\hspace*{1cm}
         +7521.91(19) N^2
         +35180.6(6) N
         +42962.6(6)
         \bigg]
          \\
      &- \frac{\gtni^8}{(N+8)^8} \bigg[-\frac{N^8}{512}
          -0.113811144429731 \ N^7
          -2.23349(12) N^6
          -9.232(14) N^5
          -116.02(24) N^4 \\ &\hspace*{1cm}
          +(9073.7\pm 2.2) N^3
          +(139051\pm 11) N^2
          +(567393\pm 30) N
          +(652860\pm 29)
        \bigg]\,.
\end{align*}\\
\end{footnotesize}

\section{Series coefficients in the $N$-component model with cubic anisotropy}
\label{app:CoeffCub}

In this appendix, we report the coefficients for the series of the RG functions in eqs.~\eqref{eq:betau}, \eqref{eq:betav}, \eqref{eq:eta}, and \eqref{eq:eta2}. The numerical coefficients appearing without error have been computed to a higher accuracy and have been truncated at $10^{-15}$ (for the coefficients smaller than $10^{-5}$ we kept 11 significant digits instead).

\begin{landscape}
\begin{table}[h!]
\scriptsize
  \centering
  \begin{tabular}{c|l}
    \hline\hline
    $i,j$			& $b^{(u)}_{ij}$       		    \\\hline
   $3,0$ 		& $0.273855167655838 + 0.075364028874214 N + 0.001850401592409 N^2$ 		\\
   $2,1$ 		& $0.677423248125902 + 0.027353408862255 N$ 	  \\
   $1,2$		    & $0.415456496888908 + 0.002559214833961 N$ 	\\
   $0,3$		    & $0.090448950786282$  \\\hline

   $4,0$ 		& $-0.2792572364240(2) - 0.09183374920899(9) N - 0.00545956462000(1) N^2 + 0.000023722894008 N^3$ 		\\
   $3,1$ 		& $-0.9438366234077(8) - 0.08325280699058(14) N +0.000618601760806 N^2$ 	  \\
   $2,2$		    & $-0.9649788763445(10) - 0.01246014444243(4) N$ 	\\
   $1,3$		    & $-0.4233187377151(6) - 0.001770942910797 N$ \\
   $0,4$		    & $-0.07544669182911(13)$  \\\hline

   $5,0$ 		& $0.35174494(11) + 0.13242509(5) N + 0.011322034(5) N^2 + 0.0000548336(2) N^3 + 8.6768976293 \ 10^{-7} N^4$ 		\\
   $4,1$ 		& $1.5209015(5) + 0.19450548(8) N + 0.001107859(4) N^2 + 0.00003177978918 N^3$ 		\\
   $3,2$ 		& $2.2073359(7) + 0.06533634(5) N +0.0003564920(8) N^2$ 	  \\
   $2,3$		    & $1.5315701(5) + 0.010676889(18) N$ 	\\
   $1,4$		    & $0.56035222(18) + 0.001346946(3) N$ \\
   $0,5$		    & $0.08749334(3)$  \\\hline

   $6,0$ 		& $-0.510500 0 ( 5 ) - 0.214853 2 ( 3 ) N - 0.0238395 1 ( 4 ) N^2 - 0.00050022 7 ( 4 ) N^3 + 2.01681 ( 15 )\ 10^{-6} N^4 + 4.4076813889 \ 10^{-8} N^5$\\
   $5,1$ 		& $-2.69841 5 ( 2 ) - 0.450684 7 ( 6 ) N - 0.0108216 7 ( 6 ) N^2 + 0.00005796 7 ( 3 ) N^3 + 2.0515471462 \ 10^{-6}N^4 $\\
   $4,2$ 		& $-5.11357 0 ( 4 ) - 0.267694 2 ( 6 ) N - 0.0006312 2 ( 3 ) N^2 + 0.000019413 6 ( 4 ) N^3 $\\
   $3,3$ 		& $-4.93174 9 ( 4 ) - 0.067575 9 ( 3 ) N + 0.0000282 83 ( 10 ) N^2$\\
   $2,4$ 		& $-2.75469 5 ( 3 ) - 0.009583 89 ( 11 ) N$\\
   $1,5$ 		& $-0.86229 87 ( 10 ) - 0.0018563 38 ( 19 ) N $\\
   $0,6$ 		& $-0.117951 38 ( 17 ) $\\\hline

   $7,0$ 		& $0.832458(3) + 0.3849068(17) N + 0.0516358(4) N^2 + 0.00184587(5) N^3 - 3.008(3) \ 10^{-6} N^4 + 1.1360(10) \ 10^{-7} N^5 + 2.6428442927 \ 10^{-9} N^6$ 		\\
   $6,1$ 		& $5.215437(16) + 1.062849(6) N + 0.0444430(9) N^2 - 0.00007846(6) N^3 + 4.2933(17) \ 10^{-6} N^4 + 1.4878449976 \ 10^{-7} N^5$ 		\\
   $5,2$ 		& $12.25220(4) + 0.935735(8) N + 0.0070463(7) N^2 + 0.00006361(2) N^3 + 1.7262(2)\ 10^{-6} N^4$ 		\\
   $4,3$ 		& $15.26308(5) + 0.367111(7) N + 0.0009380(3) N^2 + 8.239(5) \ 10^{-6} N^3$ 	  \\
   $3,4$		    & $11.59355(4) + 0.072378(3) N + 0.00020215(6) N^2$ 	\\
   $2,5$		    & $5.49610(2) + 0.0107550(8) N$ \\
   $1,6$		    & $1.506425(7) + 0.00241029(11) N$  \\
   $0,7$		    & $0.1831012(12)$  \\\hline\hline
  \end{tabular}
  \caption{The coefficients $b^{(u)}_{ij}$, see eq.~\eqref{eq:betau}.}
\end{table}

\begin{table}[h!]
\scriptsize
  \centering
  \begin{tabular}{c|l}
    \hline\hline
    $i,j$			& $b^{(v)}_{ij}$       		    \\\hline
   $3,0$ 		& $0.643805171872750 + 0.057412760213386 N - 0.001716196584448 N^2$ 		\\
   $2,1$ 		& $1.685330465609902 + 0.003071411401998 N$ 	  \\
   $1,2$		    & $1.313829441703564$ 	\\
   $0,3$		    & $0.351069598122462$  \\\hline

   $4,0$ 		& $-0.7670617665149(5) - 0.08905466691534(11) N + 0.000040711389653(9) N^2 - 0.000087586116896 N^3$ 		\\
   $3,1$ 		& $-2.738584143788(2) - 0.04921887434087(17) N - 0.000026234674 N^2$ 	  \\
   $2,2$		    & $-3.347720432576(3) + 0.00754183961192(5) N$ 	\\
   $1,3$		    & $-1.8071874449658(15)$ \\
   $0,4$		    & $-0.3765268273590(3)$  \\\hline

   $5,0$ 		& $1.09653528(3) + 0.15791303(7) N + 0.002358468(4) N^2 - 0.00006147133639(3) N^3 - 5.3871244824 \ 10^{-6} N^4$ 		\\
   $4,1$ 		& $4.9865509(16) + 0.17572812(13) N - 0.0020718380(19) N^2 - 0.000019382907423 N^3$ 		\\
   $3,2$ 		& $8.364533(3) + 0.00396214(7) N + 0.000213631289238(14) N^2$ 	  \\
   $2,3$		    & $6.894605(2) - 0.023087389(13) N$ 	\\
   $1,4$		    & $2.8857933(9)$ \\
   $0,5$		    & $0.49554777(16)$  \\\hline

   $6,0$ 		& $-1.77455 64 ( 15 ) - 0.304044 2 ( 4 ) N - 0.0094339 2 ( 4 ) N^2 + 0.00006698 9 ( 2 ) N^3 - 6.5724867 38 ( 15 )\ 10^{-6} N^4 - 3.7531135139\ 10^{-7} N^5$\\
   $5,1$ 		& $-9.82984 8 ( 7 ) - 0.53385 26 ( 11 ) N + 0.0022031 7 ( 6 ) N^2 - 0.000130668 5 ( 8 ) N^3 - 2.5959418019\ 10^{-6} N^4 $\\
   $4,2$ 		& $-21.0735 81 ( 16 ) - 0.16629 03 ( 12 ) N - 0.0000148 6 ( 2 ) N^2 + 4.4988692 3 ( 3 )\ 10^{-6} N^3 $\\
   $3,3$ 		& $-23.5697 76 ( 18 ) + 0.095715 3 ( 6 ) N - 0.00083903 8 ( 3 ) N^2 $\\
   $2,4$ 		& $-14.9280 33 ( 12 ) + 0.048681 04 ( 12 ) N$\\
   $1,5$ 		& $-5.12988 4 ( 5 ) $\\
   $0,6$ 		& $-0.749690 8 ( 9 )$\\\hline

   $7,0$ 		& $3.182705(9) + 0.632210(3) N + 0.0291750(4) N^2 + 4.94(4)\ 10^{-6} N^3 - 7.708(16)\ 10^{-7} N^4 - 6.21499283(8)\ 10^{-7} N^5 - 2.8452078911 \ 10^{-8} N^6$ 		\\
   $6,1$ 		& $20.82993(5) + 1.552103(11) N + 0.0065542(10) N^2 - 7.92(4)\ 10^{-6} N^3 - 0.0000109597(3) N^4 - 2.8940373755 \ 10^{-7} N^5$ 		\\
   $5,2$ 		& $54.54539(13) + 0.995474(17) N - 0.0047332(8) N^2 - 0.000030141(11) N^3 - 5.95573536(14)\ 10^{-7} N^4$ 		\\
   $4,3$ 		& $77.4292(2) - 0.139913(14) N + 0.0036756(3) N^2 - 0.0000420152(9) N^3$ 	  \\
   $3,4$	    & $66.01628(18) - 0.357688(6) N + 0.00179214(4) N^2$ 	\\
   $2,5$	    & $34.17580(10) - 0.1010148(11) N$ \\
   $1,6$		    & $9.98365(4)$  \\
   $0,7$	    & $1.270843(6)$  \\\hline\hline
  \end{tabular}
  \caption{The coefficients $b^{(v)}_{ij}$, see eq.~\eqref{eq:betav}.}
\end{table}

\begin{table}[h!]
\scriptsize
  \centering
  \begin{tabular}{c|l}
    \hline\hline
   $i,j$			& $e^{(\phi)}_{ij}$       		    \\\hline
   $3,0$ 		& $0.643805171872750 + 0.000338600835747 N + 0.000033860083575 N^2$ 		\\
   $2,1$ 		& $0.002437926017376 + 0.000304740752172 N$ 	  \\
   $1,2$		    & $0.002742666769548$ 	\\
   $0,3$		& $0.000914222256516$  \\\hline

   $4,0$ 		& $0.00099254837122(3) + 0.00070251806232(2) N + 0.000101811600498(3) N^2 - 0.000087586116896 N^3$ 		\\
   $3,1$ 		& $0.00595529022731(16) + 0.00123746326024(4) N - 7.8620271300 \ 10^{-6} N^2$ 	  \\
   $2,2$		& $0.0104656702615(3) + 0.000311666929101(11) N$ 	\\
   $1,3$		    & $0.0071848914604(2)$ \\
   $0,4$		& $0.00179622286510(5)$  \\\hline

   $5,0$ 		& $-0.000366591(2) - 0.0002572073(15) N - 0.0000320259(2) N^2 + 2.2430705244(14)\ 10^{-6} N^3 - 1.1094045683 \ 10^{-7} N^4$ 		\\
   $4,1$ 		& $-0.002749435(15) - 0.000554337(4) N + 0.000036974271571(16) N^2 - 1.6641068524 \ 10^{-6} N^3$ 		\\
   $3,2$ 		& $-0.00646950(4) - 0.0000702065(14) N + 2.782388315(3)\ 10^{-6} N^2$ 	  \\
   $2,3$		& $-0.00660452(4) + 0.00006759334592(4) N$ 	\\
   $1,4$		    & $-0.003268462(19)$ \\
   $0,5$		& $-0.000653692(4)$  \\\hline

   $6,0$ 		& $0.000695678(11) + 0.000565857(9) N + 0.000120572(2) N^2 + 5.74661(19)\ 10^{-6} N^3 - 3.83849767(6)\ 10^{-8} N^4 - 1.0441273326 \ 10^{-8} N^5$ 		\\
   $5,1$ 		& $0.00626110(8) + 0.00196216(3) N + 0.000104069(3) N^2 - 3.15043741(8)\ 10^{-7} N^3 - 1.8794291986 \ 10^{-7} N^4$ 		\\
   $4,2$ 		& $0.0189571(2) + 0.00184829(4) N + 0.0000124821(8) N^2 - 7.553776112(8)\ 10^{-7} N^3$ 		\\
   $3,3$ 		& $0.0271587(3)+ 0.000595692(18) N + 1.704347526(11)\ 10^{-6} N^2$ 	  \\
   $2,4$		& $0.0207756(3) + 0.000041476(4) N$ 	\\
   $1,5$		    & $0.00832682(12)$ \\
   $0,6$	    & $0.00138780(3)$  \\\hline

   $7,0$	    & $-0.00083387 ( 4 ) - 0.00069947 ( 4 ) N - 0.000157180 ( 11 ) N^{2} - 7.3827 ( 15 )\ 10^{-6} N^{3} +2.6562 ( 9 )\ 10^{-7} N^{4} - 1.27960693 ( 3 )\ 10^{-8} N^{5} - 9.1800947398 \ 10^{-10} N^{6} $\\
   $6,1$	    & $-0.0087557 ( 4 ) - 0.00296656 ( 15 ) N - 0.00016711 ( 2 ) N^{2} + 6.0384 ( 15 )\ 10^{-6} N^{3} - 2.30161056 ( 4 )\ 10^{-7} N^{4} - 1.9278198954 \ 10^{-8} N^{5}$\\
   $5,2$	    & $-0.0320491 ( 12 ) - 0.0036280 ( 3 ) N +0.000027239 ( 19 ) N^{2} - 7.137 ( 3 )\ 10^{-7} N^{3} - 1.352343475 ( 3 )\ 10^{-7} N^{4}$\\
   $4,3$	    & $-0.057921 ( 2 ) - 0.0015160 ( 2 ) N + 0.000019338 ( 7 ) N^{2} - 3.08920620 ( 4 )\ 10^{-7} N^{3}$\\
   $3,4$	    & $-0.059380 ( 2 ) - 0.00003885 ( 9 ) N + 5.116 ( 9 )\ 10^{-7} N^{2}$\\
   $2,5$	    & $-0.0357410 ( 15 ) +0.000090327 ( 18 ) N $\\
   $1,6$	    & $-0.0118836 ( 6 ) $\\
   $0,7$	    & $-0.00169765 ( 10 ) $\\\hline

   $8,0$ 		& $0.0013543(7) + 0.0012112(6) N + 0.0003177(2) N^2 + 0.00002581(4) N^3 + 2.70(4)\ 10^{-7} N^4 + 3.50(16)\ 10^{-9} N^5 - 1.6361696(4) \ 10^{-9} N^6 - 8.0689474041 \ 10^{-11} N^7$ 		\\
   $7,1$ 		& $0.016251(6) + 0.006409(3) N + 0.0006072(6) N^2 + 6.16(6)\ 10^{-6} N^3 + 1.55(3)\ 10^{-7} N^4 - 3.5394976(6)\ 10^{-8} N^5 - 1.9365473771 \ 10^{-9} N^6$ 		\\
   $6,2$ 		& $0.07018(2) + 0.011035(6) N + 0.0002457(7) N^2 + 2.32(4)\ 10^{-6} N^3 - 2.542(4)\ 10^{-7} N^4 - 1.79734602(4) \ 10^{-8} N^5$ 		\\
   $5,3$ 		& $0.15474(4) + 0.008152(7) N + 0.0000254(4) N^2 + 1.26(9)\ 10^{-7} N^3 -8.0385997(6)\ 10^{-8} N^4$ 		\\
   $4,4$ 		& $0.20121(6) + 0.002430(5) N + 9.58(14)\ 10^{-6} N^2 - 1.721(9)\ 10^{-7} N^3$ 	  \\
   $3,5$	    & $0.16299(5) - 0.000076(2) N + 4.878(18)\ 10^{-6} N^2$ 	\\
   $2,6$	    & $0.08160(3) - 0.0001403(3) N$ \\
   $1,7$		& $0.023274(10)$  \\
   $0,8$	    & $0.0029093(17)$  \\\hline\hline
  \end{tabular}
  \caption{The coefficients $e^{(\phi)}_{ij}$, see eq.~\eqref{eq:eta}.}
\end{table}

\begin{table}[h!]
\scriptsize
  \centering
  \begin{tabular}{c|l}
    \hline\hline
   $i,j$			& $e^{(\phi^2)}_{ij}$       		    \\\hline
   $3,0$ 		& $-0.025120498969573 - 0.016979919272964 N - 0.002209834894089 N^2$ 		\\
   $2,1$ 		& $-0.113042245363077 - 0.019888514046799 N$ 	  \\
   $1,2$		    & $-0.130371544575914 - 0.002559214833962 N$ 	\\
   $0,3$		& $-0.044310253136625$  \\\hline

   $4,0$ 		& $0.021460047441142 + 0.015690832591609 N + 0.002405927309779 N^2 - 0.000037238562870 N^3$ 		\\
   $3,1$ 		& $0.128760284646854 + 0.029764853226225 N - 0.000446862754441 N^2$ 	  \\
   $2,2$		& $0.227791774941802 + 0.009325637736155 N$ 	\\
   $1,3$		    & $0.156307332207841 + 0.001770942910797 N$ \\
   $0,4$		& $0.039519568779659$  \\\hline

   $5,0$ 		& $-0.022694306(17) -0.017985182(12) N -0.003583541(2) N^2 -0.00013566164(6) N^3 - 1.6993090378 \ 10^{-6} N^4$ 		\\
   $4,1$ 		& $-0.17020729(13) -0.04978522(3) N -0.0019839453(10) N^2 -0.000025489635567 N^3$ 		\\
   $3,2$ 		& $-0.4091761(3) -0.034538392(14) N -0.0002894318(2) N^2$ 	  \\
   $2,3$		& $-0.4346482(3) -0.009355701(5) N$ 	\\
   $1,4$		    & $-0.22065500(15) - 0.0013469481(10) N$ \\
   $0,5$		& $-0.04440039(3)$  \\\hline

   $6,0$ 		& $ 0.0294507 2 ( 6 ) + 0.0248746 7 ( 5 ) N + 0.0057284 21 ( 14 ) N^2 + 0.00031558 06 ( 14 ) N^3 - 5.8586 8 ( 4 )\ 10^{-6} N^4 - 1.0373505488\ 10^{-7} N^5$\\
   $5,1$ 		& $ 0.265056 4 ( 5 ) + 0.091343 8 ( 2 ) N + 0.0058839 0 ( 2 ) N^2 - 0.000101721 9 ( 5 ) N^3 - 1.8672309881\ 10^{-6} N^4$\\
   $4,2$ 		& $ 0.80694 93 ( 15 ) + 0.097982 9 ( 3 ) N + 0.00053193 5 ( 8 ) N^2 - 0.0000128197 4 ( 9 ) N^3 $\\
   $3,3$ 		& $ 1.16381 5 ( 2 ) + 0.043471 28 ( 15 ) N - 0.00001786 68 ( 16 ) N^2 $\\
   $2,4$ 		& $ 0.89481 70 ( 17 ) + 0.0106342 9 ( 4 ) N $\\
   $1,5$ 		& $ 0.360324 2 ( 7 ) + 0.00185633 7 ( 6 ) N $\\
   $0,6$ 		& $ 0.060363 42 ( 14 ) $\\\hline

   $7,0$ 		& $-0.0435337(7) -0.0388023(7) N -0.0101019(2) N^2 -0.000806066(33) N^3 - 7.691(2)\ 10^{-6} N^4 - 3.6317(2)\ 10^{-7}N^5 - 7.2733914189\ 10^{-9} N^6$ 		\\
   $6,1$ 		& $-0.457104(6) -0.178872(3) N -0.0166337(5) N^2 -0.00014687(4) N^3 - 7.3210(3)\ 10^{-6} N^4 - 1.5274121980\ 10^{-7} N^5$ 		\\
   $5,2$ 		& $-1.68697(2) -0.265203(5) N -0.0060545(5) N^2 -0.000061887(9) N^3 -1.20284(4)\ 10^{-6} N^4$ 		\\
   $4,3$ 		& $-3.08675(4) -0.176101(5) N -0.0009588(2) N^2 -5.8441(8)\ 10^{-6} N^3$ 	  \\
   $3,4$	    & $-3.20124(4) -0.062512(3) N -0.00006546(4) N^2$ 	\\
   $2,5$	    & $-1.94400(3) -0.0142947(7) N$ \\
   $1,6$		    & $-0.650354(10) -0.00241008(9) N$  \\
   $0,7$	    & $-0.0932520(17)$  \\\hline

    $8,0$ 		& $0.072097(8) + 0.067415(8) N + 0.019369(3) N^2 + 0.0019259(6) N^3 + 0.00004027(7) N^4 - 6.98(4)\ 10^{-7} N^5 - 2.678(3)\ 10^{-8} N^6 - 5.5497030817\ 10^{-10} N^7$ 		\\
    $7,1$ 		& $0.86516(7) + 0.37640(4) N + 0.044227(9) N^2 + 0.0009974(12) N^3 - 0.00001553(7) N^4 - 6.160(6) \ 10^{-7} N^5 - 1.3319287575\ 10^{-8} N^6$ 		\\
    $6,2$ 		& $3.7560(3) + 0.72036(9) N + 0.027371(13) N^2 - 0.0000327(9) N^3 - 5.784(14)\ 10^{-6} N^4 - 1.2476(5) \ 10^{-7} N^5$ 		\\
   $5,3$ 		& $8.3459(6) + 0.65528(11) N + 0.006210(9) N^2 - 0.0000343(3) N^3 - 7.096(11) \ 10^{-7} N^4$ 		\\
   $4,4$ 		& $10.9313(7) + 0.32736(8) N + 0.000556(4) N^2 - 2.60(4)\ 10^{-6} N^3$ 		\\
   $3,5$ 		& $8.9075(7) + 0.09978(4) N + 0.0000380(7) N^2$ 	  \\
   $2,6$		& $4.4807(4) + 0.023020(11) N$ 	\\
   $1,7$		& $1.28285(13) + 0.0039158(15) N$ \\
   $0,8$	    & $0.16085(2)$  \\\hline\hline
  \end{tabular}
  \caption{The coefficients $e^{(\phi^2)}_{ij}$, see eq.~\eqref{eq:eta2}.}
\end{table}

\end{landscape}

\FloatBarrier

\bibliographystyle{JHEP}
\bibliography{refs}

\end{document}